\documentclass[sigconf]{acmart}
\AtBeginDocument{%
  }

\setcopyright{none}
\renewcommand\footnotetextcopyrightpermission[1]{}
\acmConference[]{}{}{}
\usepackage[textsize=tiny]{todonotes}
\usepackage{colortbl}
\usepackage{tcolorbox }
\usepackage{algorithm}
\usepackage{array}

\usepackage{algorithmic}
\definecolor{bluei}{RGB}{218,232,252}
\definecolor{nmgray}{RGB}{229,229,229}

\newcommand{\sjcha}[1]{\textcolor{black}{#1}}

\newtcolorbox{mybox}[2][]{
width=\columnwidth,
colback = nmgray!75!white, 
colframe = nmgray!75!white, 
boxsep=0pt,left=5pt,right=20pt,top=10pt,bottom=10pt,
fontupper=\linespread{0.9}\selectfont,
title=#2,#1}
\begin{document}

\title{CatchPhrase: EXPrompt-Guided Encoder Adaptation for Audio-to-Image Generation}
\author{Hyunwoo Oh}
\email{komjii@hanyang.ac.kr}
\affiliation{%
  \institution{Hanyang University}
  \city{Seoul}
  \country{Republic of Korea}
}

\author{SeungJu Cha}
\email{sju9020@hanyang.ac.kr}
\affiliation{%
  \institution{Hanyang University}
  \city{Seoul}
  \country{Republic of Korea}
}

\author{Kwanyoung Lee}
\email{mobled37@hanyang.ac.kr}
\affiliation{%
  \institution{Hanyang University}
  \city{Seoul}
  \country{Republic of Korea}
}

\author{Si-Woo Kim}
\email{boreng0817@hanyang.ac.kr}
\affiliation{%
 \institution{Hanyang University}
  \city{Seoul}
  \country{Republic of Korea}
}
\author{Dong-Jin Kim}
\authornote{Corresponding author.}
\email{djdkim@hanyang.ac.kr}
\affiliation{%
  \institution{Hanyang University}
  \city{Seoul}
  \country{Republic of Korea}
}

\renewcommand{\shortauthors}{}

\begin{abstract}
We propose \textbf{CatchPhrase}, a novel audio-to-image generation framework designed to mitigate semantic misalignment between audio inputs and generated images. 
While recent advances in multi-modal encoders have enabled progress in cross-modal generation, ambiguity stemming from \textit{homographs} and \textit{auditory illusions} continues to hinder accurate alignment.
To address this issue, CatchPhrase generates enriched cross-modal semantic prompts (\textbf{EXPrompt Mining}) from weak class labels by leveraging large language models (LLMs) and audio captioning models (ACMs). 
To address both class-level and instance-level misalignment, we apply multi-modal filtering and retrieval to select the most semantically aligned prompt for each audio sample (\textbf{EXPrompt Selector}). 
A lightweight mapping network is then trained to adapt pre-trained text-to-image generation models to audio input. 
Extensive experiments on multiple audio classification datasets demonstrate that CatchPhrase improves audio-to-image alignment and consistently enhances generation quality by mitigating semantic misalignment.
\footnote{Project page: https://github.com/komjii2/CatchPhrase}

\end{abstract}



\begin{teaserfigure}
  \centering
  \includegraphics[width=\textwidth]{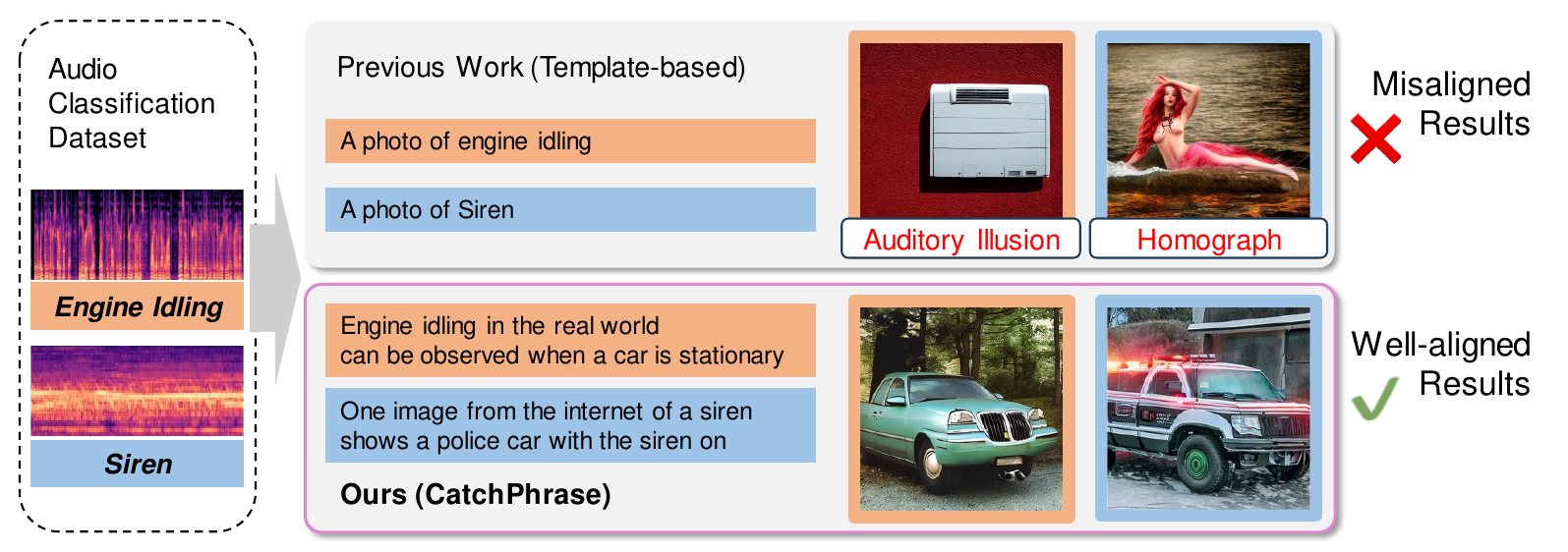}
  \caption{Visual demonstration of CatchPhrase. \textmd{
  Unlike prior method~\cite{qin2023gluegen} relying on fixed templates for audio-text alignment, CatchPhrase uses semantically enriched prompts from audio and text cues, producing well-aligned results for each audio sample.}}
  \label{fig:teaser}
\end{teaserfigure}

\maketitle

\section{Introduction}
\begin{figure}
    \centering
    \includegraphics[width=1.0\linewidth]{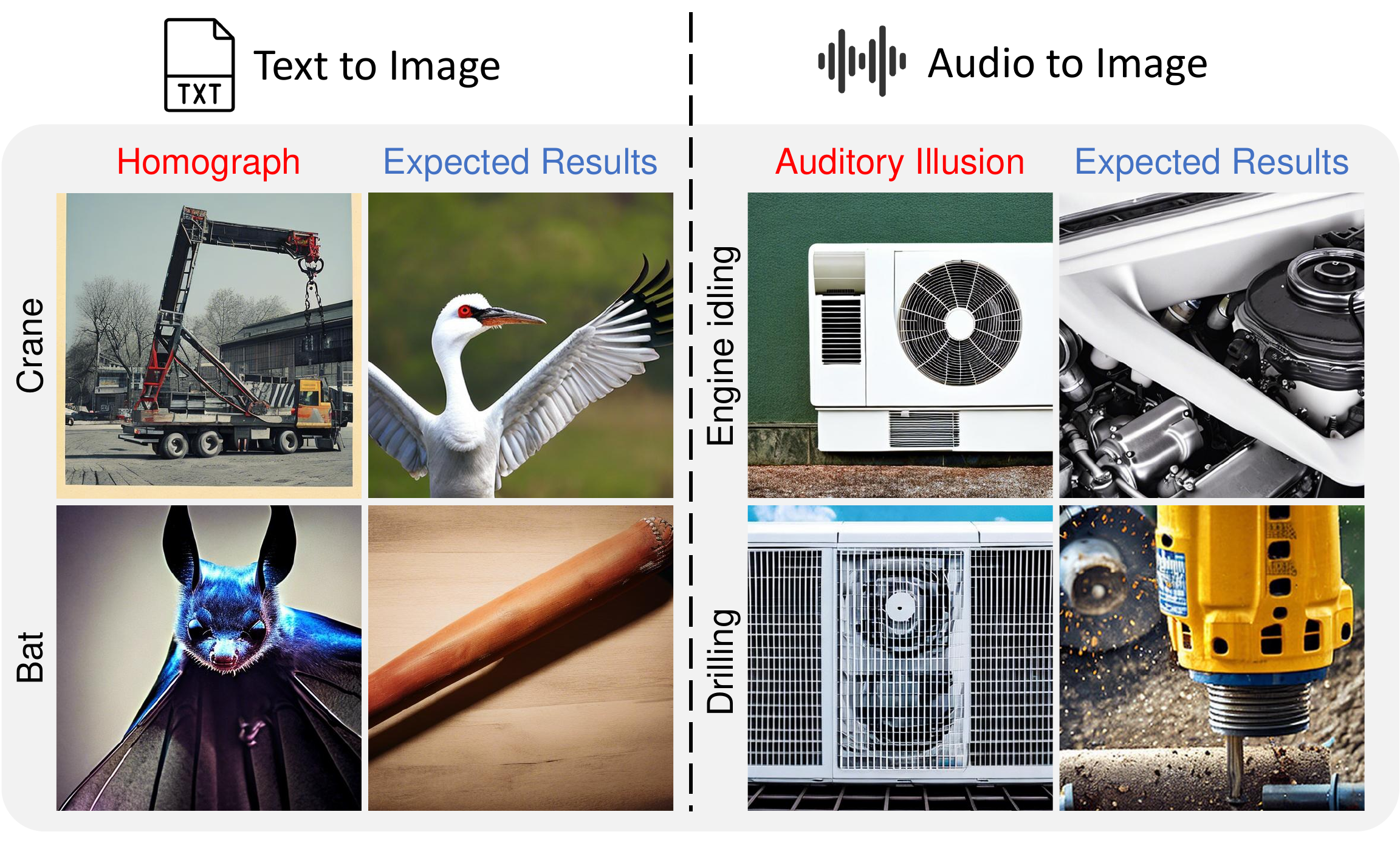}
    \caption{\textbf{Examples of cross-modal misalignment.} \textmd{When the prompts ``crane'' and ``bat'' are used for image generation, the model often confuses their meanings, resulting in class-level misalignment (i.e., homographs).
    Similarly, given audio samples such as ``engine idling'' or ``drilling,'' the model produces instance-level misaligned results due to auditory illusions.}}
    \label{fig:Cross-modal Ambiguity}
\end{figure}

With the recent advances in text-to-image generation~\cite{nichol2021glide, ramesh2022hierarchical, rombach2022high, saharia2022photorealistic}, there has been growing interest in extending pre-trained models by adding new modalities without end-to-end retraining~\cite{li2023gligen, qin2023gluegen, zhang2023adding, biner2024sonicdiffusion}.
\sjcha{Among these, audio-to-image generation~\cite{sung2023sound, yariv2023audiotokenadaptationtextconditioneddiffusion, lee2023generatingrealisticimagesinthewild, biner2024sonicdiffusion} has gained attention for its potential for cross-modal generation, enabled by advances in multi-modal encoders~\cite{guzhov2022audioclip, wu2023large, girdhar2023imagebind}.}
To enable audio-based conditioning, recent work~\cite{qin2023gluegen, yariv2023audiotokenadaptationtextconditioneddiffusion} proposed transferring knowledge from a text encoder to an audio encoder using an audio-text paired dataset.
\sjcha{However, due to the scarcity and difficulty of collecting detailed captions for audio,}
GlueGen~\cite{qin2023gluegen} instead utilized audio classification datasets (i.e., audio-category pairs) by combining each class label with simple CLIP~\cite{radford2021learning}-style prompt templates.

\sjcha{Although GlueGen~\cite{qin2023gluegen} showed notable performance in audio-to-image generation, we observe that relying on a simple fixed template introduces two types of semantic misalignment: a class-level linguistic misalignment and an instance-level auditory misalignment, as illustrated in Fig.~\ref{fig:Cross-modal Ambiguity}.}
First, the linguistic misalignment arises from phenomena such as \textit{homographs} (or polysemy), a problem previously identified in text-to-image generation~\cite{du2024stable}.
As shown in the left part of Fig.~\ref{fig:Cross-modal Ambiguity}, the model sometimes confuses words like ``crane'' and ``bat,'' leading to mismatched image outputs.
This occurs because audio classification datasets consist of simple labels that lack sufficient semantic information to disambiguate between identical words with different meanings.
Second, at the instance level, auditory misalignment like \textit{auditory illusion} arises due to limited semantic cues in audio signals, making classification less reliable.
For instance, even when input sounds are clearly distinguishable to humans, the model often fails to capture such perceptual differences, as shown in the right part of Fig.~\ref{fig:Cross-modal Ambiguity}.
While auditory misalignment is widely observed in diverse perceptual phenomena, our focus is on these general-context misalignments, rather than phoneme-level effects such as the McGurk effect~\cite{mcgurk1976hearing}, which involve mismatches between auditory and visual speech cues.

Based on these observations, we hypothesize that enriching each class with appropriate semantic information can help reduce the inherent misalignment arising from the limited expressiveness of class labels in audio classification datasets.
To improve semantic alignment in audio-to-image generation, we propose \textbf{CatchPhrase}, which utilizes audio-informed and context-rich prompts to mitigate cross-modal misalignment.
\sjcha{However, crafting such prompts with diverse contextual details for every class is labor-intensive.}
To automate the generation of meaningful, audio-informed prompts from weak labels, we extract Enriched Cross-modal Prompts (EXPrompt), leveraging class-related knowledge from large language models (LLMs)~\cite{radford2018improving,radford2019language,ouyang2022traininglanguagemodelsfollow} and audio captioning models (ACMs)~\cite{kim2024enclap}.
Specifically, we construct semantically enriched text prompts by querying LLMs with inputs composed of three perspectives (visual, auditory, and semantic) combined with weak class labels.
To further incorporate contextual cues from the audio itself, we add enriched audio prompts using zero-shot ACMs.
EXPrompts capture a wide range of class-relevant contextual information.

To enhance semantic alignment and assign appropriate prompts to each audio input, we introduce EXPrompt Selector, a multi-modal aware filtering and retrieval strategy. 
We first compute the relevance scores between EXPrompts and an audio subset within the same class from the audio classification dataset, filtering out class-irrelevant prompts. 
This process mitigates class-level misalignment, such as confusion cased by homographs, and ensures that each audio class is aligned with contextually appropriate prompts.
To further address instance-level misalignment arising from perceptual discrepancies, such as auditory illusions, we apply a retrieval step. 
For each audio sample, we compute its similarity with the filtered prompt set within the same class and retrieve the prompt that is most semantically aligned.
This enables fine-grained alignment by reflecting subtle variations across audio samples within the same class and pairs each audio input with a prompt that best reflects its semantic context.
By grounding each audio input in a more accurate textual description, the generation model becomes more robust to perceptual inconsistencies.

Finally, we train our model using the InfoNCE loss, encouraging the generation model to produce semantically aligned images when given audio inputs while mitigating cross-modal misalignment.
We validate the effectiveness of our method on various audio classification datasets, including UrbanSound8K (US8K) ~\cite{salamon2014dataset}, ESC50 ~\cite{piczak2015esc}, and VGGSound ~\cite{chen2020vggsound}. 
Our model successfully captures semantic cues from both text and audio, effectively mitigating cross-modal misalignment, such as homographs and auditory illusions, in extensive experiments.
In conclusion, our contributions are as follows: 
\begin{itemize}
\item We propose \textbf{CatchPhrase}, an audio-to-image generation framework that explicitly addresses cross-modal misalignment. To the best of our knowledge, this is the first work to tackle auditory illusions directly in this context.
\item We introduce EXPrompts, enriched prompts that compensate for the lack of semantic detail in conventional audio classification labels.
\item We design a multi-modal-aware selector that pairs each audio sample with a semantically precise and contextually aligned prompt, improving semantic alignment during training.
\item Extensive experiments on multiple audio classification datasets validate the effectiveness of our method in reducing cross-modal misalignment.
\end{itemize}

%\djkim{======------checked roughly up to here-----======}

\section{Related Work}
%Too long
%Need to shorten

\subsection{LLMs Descriptions For Downstream Task}
Recently, with the remarkable performance of Large Language Models (LLMs) such as GPT \cite{brown2020language, achiam2023gpt} in the field of language understanding, many researchers have leveraged LLMs for various vision and language tasks \cite{menon2022visualclassificationdescriptionlarge, rotstein2023fusecap, li2024learning, hu2022promptcap, pratt2023does, abdelhamed2024you, chiquier2025evolving}.
CuPL \cite{pratt2023does} extracts appearance descriptions of each class through GPT-3 and utilizes them for better zero-shot image classification. Further, \cite{abdelhamed2024you} enhances zero-shot classification performance by leveraging enriched descriptions not only from text but also from visual inputs.
In image generation fields, \cite{hao2024optimizing} optimizes text prompts using LLMs, guiding Stable Diffusion \cite{rombach2022high} to generate better-quality images.
Instructpix2pix \cite{brooks2023instructpix2pix} and WavCaps \cite{mei2024wavcaps} create a dataset using descriptions from GPT \cite{brown2020language}, enabling image editing through human instructions. 
% Additionally, \cite{kwon2024zero} extracts global and local descriptions through LLMs and applies them to the image outpainting task.
Audio Journey\cite{michaels2024audio} leverages enriched captions from Alpaca \cite{alpaca} to augment weakly labeled audio datasets in text-to-audio generation tasks. 
Our work builds on this approach but differs by leveraging multi-modal cues to generate semantically enriched prompts.% We leverage this approach to generate rich prompts from simple class label in our study.

\subsection{Audio-to Image Generation}
% With the advancements in multi-modal learning enable the alignment of diverse modalities such as text, audio, and image within a shared embedding space.
\sjcha{Advancements in multi-modal learning have enabled the alignment of diverse modalities such as text, audio, and image within a shared embedding space.}
CLIP~\cite{radford2021learning} jointly embeds text and image, supporting a wide range of downstream tasks, including conditional image generation~\cite{rombach2022high,saharia2022photorealistic} and cross-modal retrieval~\cite{zhang2023remodiffuse}. 
Several approaches extend this representation space to include audio ~\cite{wu2022wav2clip, guzhov2022audioclip, elizalde2023clap, girdhar2023imagebind}, training audio encoders to align with the vision-language space.
% , such as Wav2CLIP~\cite{wu2022wav2clip}, AudioCLIP~\cite{guzhov2022audioclip}, CLAP~\cite{elizalde2023clap} and ImageBind~\cite{girdhar2023imagebind}, which train audio encoders to align with the vision-language space. 
% These unified representations also facilitate conditional image generation from audio.
\sjcha{These unified representations also facilitate conditional image generation from audio, which has led to significant advances in audio-to-image generation ~\cite{lee2022sound, yariv2023audiotokenadaptationtextconditioneddiffusion, qin2023gluegen, biner2024sonicdiffusion, sungbin2024soundbrushsoundbrushvisual}.}
% ~\cite{lee2022sound, yariv2023audiotokenadaptationtextconditioneddiffusion}, AudioToken~\cite{yariv2023audiotokenadaptationtextconditioneddiffusion}, GlueGen~\cite{qin2023gluegen} utilizes AudioCLIP ~\cite{guzhov2022audioclip} to enable editing or generating the image from the shared latent space.
Some methods use CLAP~\cite{elizalde2023clap} for audio-conditioned image generation~\cite{biner2024sonicdiffusion, sungbin2024soundbrushsoundbrushvisual}, while others rely on AudioCLIP~\cite{guzhov2022audioclip} to achieve similar goals~\cite{lee2022sound, yariv2023audiotokenadaptationtextconditioneddiffusion, qin2023gluegen}.
In this work, we build upon these multi-modal encoders for audio-to-image generation and focus on improving their capability to resolve cross-modal ambiguities.

\subsection{Enhancing Semantic Alignment}

In various generative tasks, including captioning and image synthesis, many studies have focused on improving semantic alignment between visual content and language.  
For example, prior work~\cite{rassin2023linguistic,zawar2024diffusionpidinterpretingdiffusionpartial,oshima2025inferencetimetexttovideoalignmentdiffusion, Kim2025SIDA, Kim2025sync} addresses the limitations of diffusion models in accurately reflecting textual prompts, enhancing text-image correspondence through attention alignment, interpretability, or inference-time guidance.
Additionally, studies such as~\cite{momeni2023verbsactionimprovingverb, cha2025verbdifftextonlydiffusionmodels} highlight that verb semantics—especially those involving human-object interactions—are often underrepresented in generated images.  
To address this, they propose training strategies that explicitly account for verb-specific structures.
In the video domain, AADiff~\cite{lee2023aadiffaudioalignedvideosynthesis} tackles temporal ambiguity by incorporating auxiliary audio signals to disambiguate motion, demonstrating the effectiveness of multimodal cues in resolving under-specified linguistic inputs.

Inspired by these insights, our approach leverages both audio and text as complementary modalities to disambiguate audio-to-image generation and improve cross-modal semantic consistency.

\begin{figure*}[t]
    \centering
    \includegraphics[width=.9\linewidth]{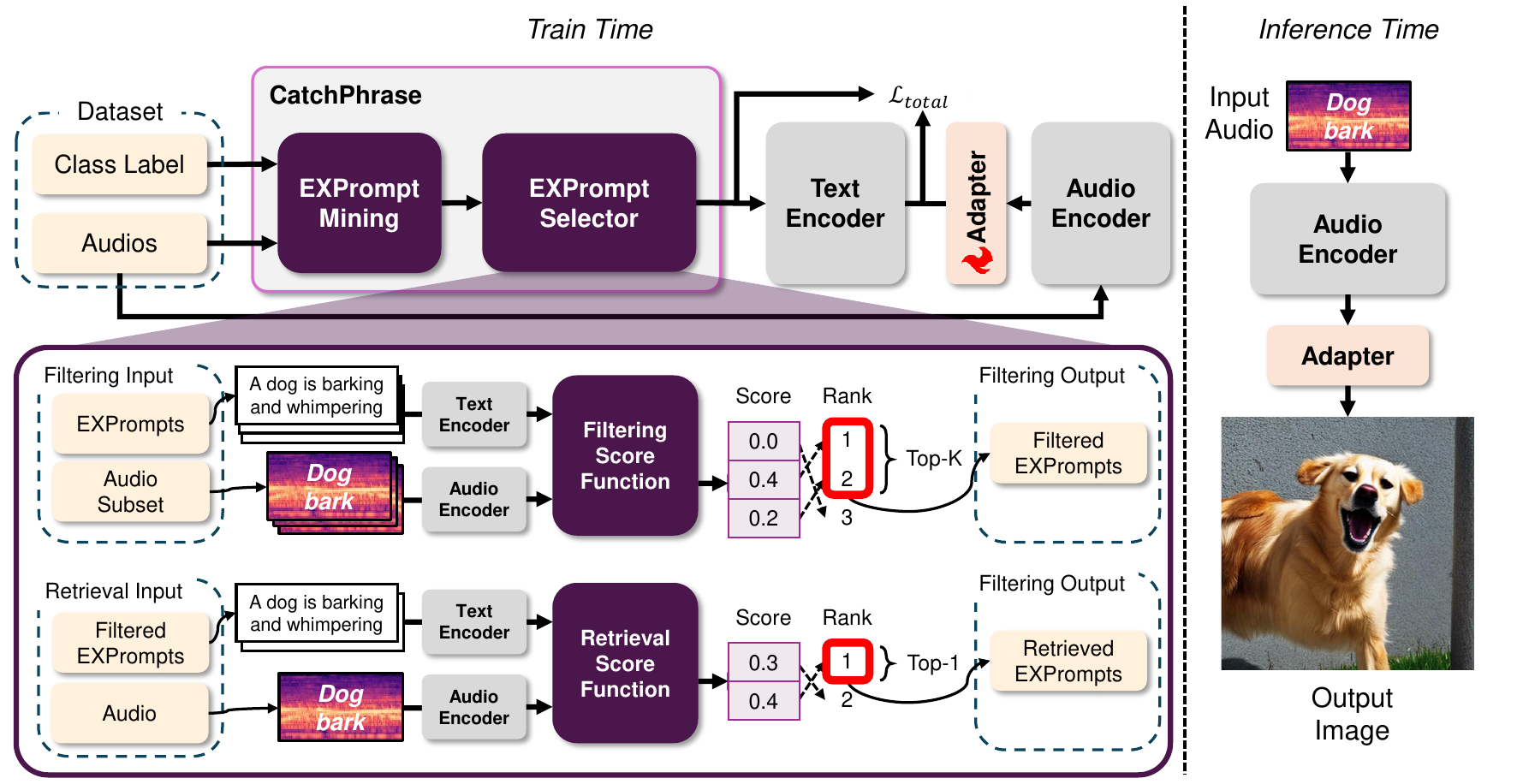} 
    \caption{\textbf{Pipeline of each stage in CatchPhrase.} 
    \textmd{(Left) EXPrompts are generated by leveraging class-relevant knowledge from LLMs and ACMs. To pair each audio sample with the most semantically appropriate prompt, we introduce the EXPrompt Selector, which enhances both class-level and instance-level alignment during adpater training. (Right) Illustration of the inference-time pipeline.}
    }
    \label{fig:CatchPhrase-Method-Diagram}
\end{figure*}

\section{CatchPhrase}
\sjcha{
Our goal is to help audio-to-image generation models mitigate cross-modal misalignment, such as homographs and auditory illusions.}
Our proposed method, \textbf{CatchPhrase}, illustrated in Fig.~\ref{fig:CatchPhrase-Method-Diagram}, consists of three main components.
First, we generate Enriched Cross-modal Prompts (EXPrompts) by leveraging class-relevant knowledge from large language models (LLMs) and audio captioning models (ACMs).
These prompts enrich the weak labels in audio classification datasets by capturing diverse and detailed semantic information. 
Second, we introduce the EXPrompt Selector, a multi-modal-aware filtering and retrieval strategy designed to reduce cross-modal misalignment. 
It aligns each audio sample with the most semantically relevant prompt by addressing both class-level and instance-level misalignment.
Finally, we train a lightweight mapping network using the enriched audio-text pairs, where the selected prompts act as semantic anchors to guide cross-modal alignment.

\subsection{Preliminary}

To condition audio inputs into pre-trained text-to-image generation model, GlueGen~\cite{qin2023gluegen} introduces a mapping network $M(\cdot)$ that aligns features from an audio encoder $AE(\cdot)$~\cite{guzhov2022audioclip} with those from a text encoder $TE(\cdot)$~\cite{radford2021learning}.
The mapping network is trained using paired text $X^{p}$ and audio $X^{a}$ samples from an audio classification dataset with three objectives.
First, features are extracted using the encoders as $f^{t} = TE(X^p)$ and $f^{a} = AE(X^a)$.
To minimize the element-wise discrepancy between $f^t$ and the mapped audio feature $M(f^a)$, they utilize a mean squared error (MSE) loss:
\begin{equation} \label{eq-mse}
\mathcal{L}_{mse} = \mathbb{E}_{(X^p, X^a)} \left[ \| f^t - M(f^a) \|_2^2 \right]
\end{equation}
In order to preserve the original information from $f^a$, they introduce a reconstruction loss using a decoder network $N(\cdot)$:
\begin{equation} \label{eq-rec}
\mathcal{L}_{rec} = \mathbb{E}_{X^a} \left[ \| f^a - N(M(f^a)) \|_2^2 \right]
\end{equation}
Finally, they apply an adversarial loss with a discriminator network $D(\cdot)$ to align the overall distribution between $f^t$ and $M(f^a)$:
\begin{equation} \label{eq-adv}
\mathcal{L}_{adv} = \mathbb{E}_{X^p} [\log D(f^t)] + \mathbb{E}_{X^a} [\log(1 - D(M(f^a)))]
\end{equation}

\subsection{EXPrompt Mining}

\begin{figure}[t]
    \centering
    \includegraphics[width=0.9\linewidth]{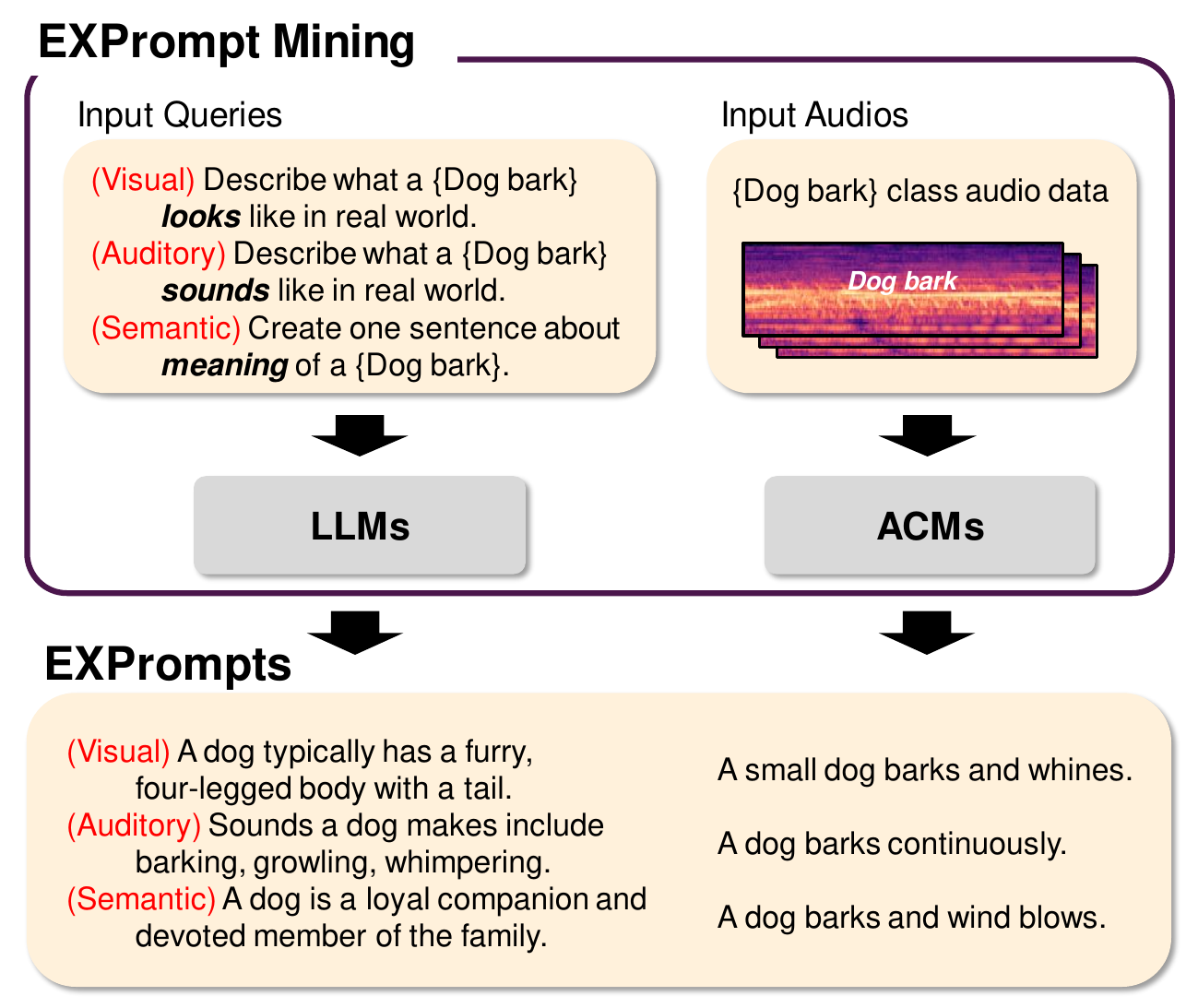} 
    \caption{\textbf{Overview of EXPrompt Mining and EXPrompt.} \textmd{We input visual, auditory, and semantic queries into LLMs and audio samples into ACMs to generate EXPrompts. While the LLM-based EXPrompts produces multiple answers per class-wise query, the ACM-based EXPrompts yields a single prompt per audio instance.}
    }
    \label{fig:EXPrompt}
\end{figure}
To enrich weak labels with semantic details in audio classification datasets, we generate semantically enriched cross-modal prompts by leveraging both textual and auditory cues, as shown in Fig.~\ref{fig:EXPrompt}.
Specifically, we extract text-informed and audio-derived rich prompts using large language models (LLMs) and audio captioning models (ACMs), respectively. 
First, we obtain class-relevant rich prompts using GPT-3.5~\cite{ouyang2022traininglanguagemodelsfollow}.
Inspired by the findings of~\cite{pratt2023does}, we design three types of queries—visual, auditory, and semantic—to capture class-relevant contextual information from LLMs.
The visual and auditory queries guide the LLM to describe observable features such as object appearance or characteristic sound patterns commonly associated with the class.
In contrast, the semantic query (e.g., ``Create one sentence about meaning of a(n) {class} in real world.'') aims to reduce linguistic ambiguity (e.g., the word ``siren'') by encouraging the generation of more specific, disambiguated, and semantically precise responses.  
The exact queries and the examples of EXPrompts are provided in the supplementary materials.

% This multi-perspective querying strategy enhances prompt diversity and .....
To additionally capture auditory cues, we generate audio-derived prompts using EnCLAP~\cite{kim2024enclap}, a zero-shot audio captioning model.
By combining prompts from both modalities, we construct a diverse and semantically rich candidate prompt set that facilitates more accurate cross-modal alignment.

\subsection{EXPrompt Selector}
%\subsection{Filtering \& Retrieval}
\sjcha{To further mitigate cross-modal misalignment that may remain in the EXPrompts,
we apply CLAP-based filtering and retrieval and pair the most semantically relevant prompts for each audio sample.
This enables the audio-to-image generation model to understand each audio sample not only at the class level but also at the instance level after training.}

\begin{figure*}[t]
    \centering
    \includegraphics[width=0.9\linewidth]{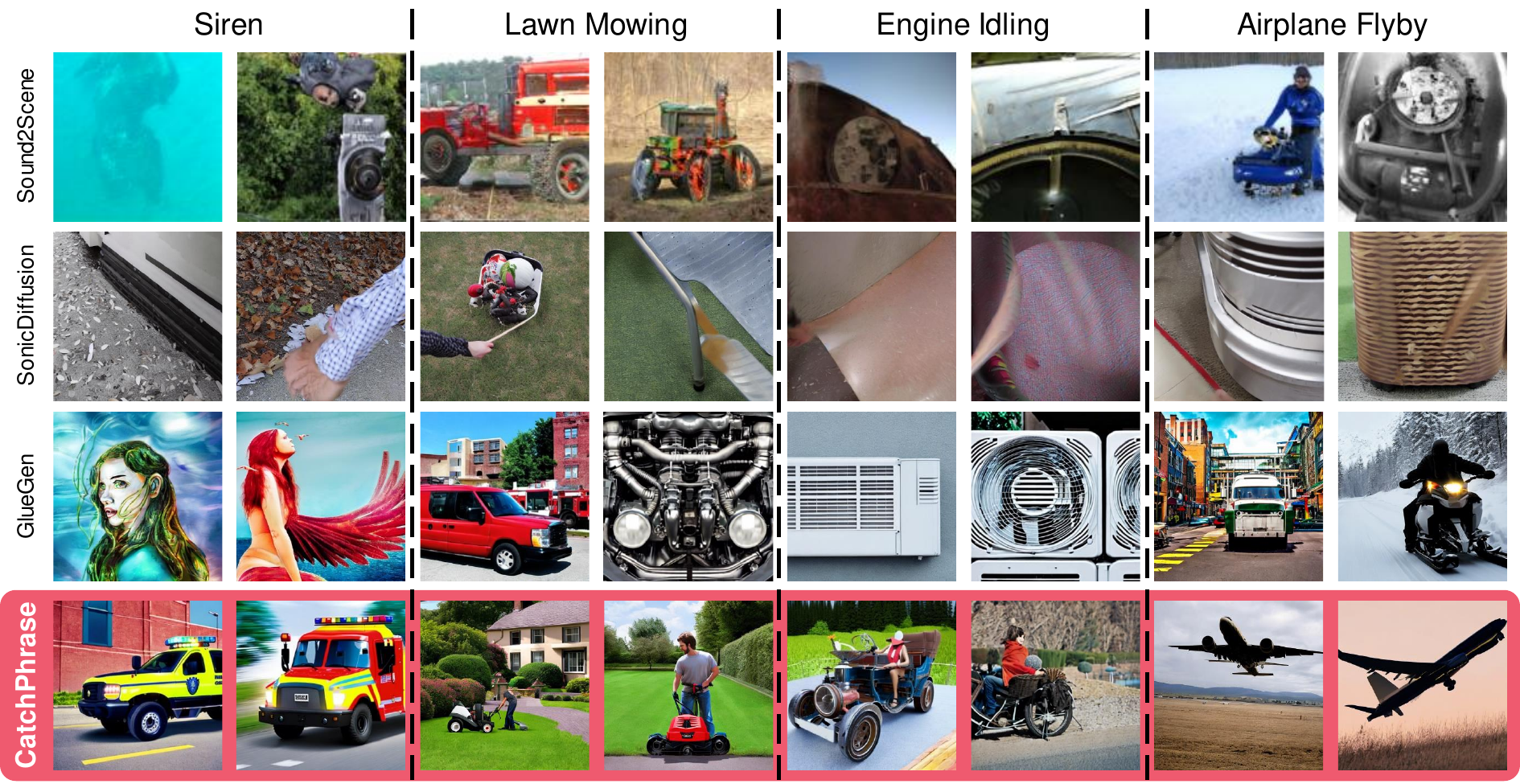}
    \caption{\textbf{Comparison of image results between previous works and CatchPhrase, demonstrating enhanced semantic alignment.} \textmd{Images with a pink borderline are generated using CatchPhrase.
    CatchPhrase generates semantically accurate images, unlike prior methods that struggle with homographs (e.g., ``siren'') and auditory illusions (e.g., ``lawn mowing,'' ``engine idling,'' 
    and ``airplane flyby'').}}
    \label{fig:Solving_Cross-modal_Ambiguity}
\end{figure*}

\noindent\textbf{EXPrompt Filtering.}
To address class-level ambiguities, particularly those arising from homographs, we apply a filtering based on the CLAP model~\cite{elizalde2023clap}, which encodes both audio and text into a shared embedding space.
For each class, we randomly sample $N_{AS}$ audio clips from the training set and compute similarity scores between each EXPrompt and the sampled audio subset. 
The relevance score function is defined as: 
\begin{equation} \label{eq-filtering}
\begin{split}
\ S_{F}(r^p) = \sum_{i\in I} (sim(r^a_i, r^p)) + \sum_{j\in J} (1 - sim(r^a_j, r^p)),\
\end{split}
\end{equation}
where $sim$ denote cosine similarity score, $r^p$ and $r^a$ denote the CLAP embeddings of the prompt and audio, respectively.  
$I$ is the set of audio samples with the same class label as the prompt, while $J$ contains audio samples from different classes.  
This scoring function penalizes prompts that are overly similar to audio samples from other classes, effectively filtering out semantically ambiguous candidates and improving class-level disambiguation.  
By selecting the top-$K$ prompts per class, we obtain a filtered EXPrompt pool that is more class-consistent and semantically disambiguated.
This helps the model resolve homograph-related confusion and class-level perceptual inconsistencies during training as shown in Fig.~\ref{fig:Solving_Cross-modal_Ambiguity}.

\begin{algorithm}[t]
\caption{: EXPrompt Selector for Filtering}
\label{alg:topk_similarity}
\begin{algorithmic}[1]
\REQUIRE Audio Subset embeddings and class label $\mathcal{A} = \{\mathbf{r}^a_i\ \text{,} \mathbf{c}_i\}_{i=1}^{N_{AS}}$, \\Text embeddings and class label $\mathcal{T} = \{\mathbf{r}^p_j\ \text{,}\mathbf{c}_j\}_{j=1}^{N_{EXP}}$, Top-$K$ parameter $K$
\ENSURE Score matrix $S_{F} \in \mathbb{R}^{N_{AS} \times N_{EXP}}$, Top-$K$ text embeddings per audio embedding

\STATE Initialize similarity matrix $S_{F} \leftarrow \text{zeros}(N_{AS}, N_{EXP})$
\FOR{$i = 1$ to $N_{AS}$}
    \STATE $\mathbf{r}^a_i, \mathbf{c}_i \leftarrow \mathcal{A}[i]$
    \FOR{$j = 1$ to $N_{EXP}$}
        \STATE $\mathbf{r}^p_j, \mathbf{c}_j\leftarrow \mathcal{T}[j]$
        \IF{$\mathbf{c}_i==\mathbf{c}_j$} 
            \STATE $S[i][j] \leftarrow S[i][j] +\text{cosine similarity}(\mathbf{r}^a_i, \mathbf{r}^p_j)$
        \ELSE
            \STATE $S[i][j] \leftarrow S[i][j] + (1-\text{cosine similarity}(\mathbf{r}^a_i, \mathbf{r}^p_j))$
        \ENDIF
        
    \ENDFOR
\STATE $I^{(i)}_{\text{top-}K} \leftarrow \text{top-K Indices}(S_{F}[i], K)$
\STATE $\mathcal{T}_{\text{top-}K}^{(i)} \leftarrow \{ \mathcal{T}[j] \mid j \in I^{(i)}_{\text{top-}K} \}$
\ENDFOR
\RETURN $S_F, \{\mathcal{T}_{\text{top-}K}^{(i)}\}_{i=1}^{N_{AS}}$
\end{algorithmic}
\end{algorithm}

\noindent\textbf{EXPrompt Retrieval.}
\sjcha{To address instance-level misalignment such as auditory illusions, we apply a retrieval step that assigns each audio sample the most semantically appropriate prompt from the filtered set.}
Given an audio embedding $r^a$ and embedding of filtered EXPrompt pool $r^p_i \in F$, we compute the cosine similarity:
\begin{equation} \label{eq-retrieval}
\ S_{R}(r^a,r^p_i)=sim(r^a, r^p_{i\in F}),\
\end{equation}
where $F$ denotes the set of filtered EXPrompts corresponding to the same class as the input audio, $N_F$ denotes number of elements in $F$.
This instance-level matching allows the model to capture subtle variations between audio inputs (e.g., different types of sirens), thereby effectively reducing perceptual ambiguity.
The retrieved EXPrompt is then paired with the corresponding audio and used to supervise the training of the mapping network.
Together, this two-stage refinement process ensures that prompts are semantically relevant and better aligned with audio inputs, thereby improving cross-modal consistency.
The overall process of the EXPrompt Selector is summarized in Alg.~\ref{alg:topk_similarity} and Alg.~\ref{alg:top1_similarity}.

\begin{algorithm}[t]
\caption{: EXPrompt Selector for Retrieval}
\label{alg:top1_similarity}
\begin{algorithmic}[1]
\REQUIRE Audio embedding $\mathbf{r}^a$, Text embeddings $\mathcal{T} = \{\mathbf{r}^p_i\}_{i=1}^{N_{F}}$
\ENSURE Score matrix $S_{R} \in \mathbb{R}^{N_{F}}$, Top-1 text embedding

\STATE Initialize similarity matrix $S_{R} \leftarrow \text{zeros}(N_{F})$
\FOR{$i = 1$ to $N_F$}
        \STATE $\mathbf{r}^p_i\leftarrow \mathcal{T}[i]$
        \STATE $S_{R}[i] \leftarrow \text{cosine similarity}(\mathbf{r}^a, \mathbf{r}^p_i)$
\ENDFOR
\STATE $I_{\text{top-}1} \leftarrow \text{top-1 Indices}(S_{R}[i], 1)$
\STATE $\mathcal{T}_{\text{top-1}} \leftarrow \{ \mathcal{T}[i] \mid i \in I_{\text{top-1}} \}$
\RETURN $S_{R}$, $\mathcal{T}_{\text{top-1}}$
\end{algorithmic}
\end{algorithm}

\subsection{Training Phase}
% \noindent\textbf{Objective Function.}

We incorporate the InfoNCE~\cite{oord2018representation} loss as an additional training objective to enhance cross-modal alignment, which is widely used for contrastive learning in representation learning frameworks. 
InfoNCE is designed to minimize the distance between a query and its corresponding positive sample while maximizing the distance to negative samples.

\begin{equation}\label{eq-infoNCE}
\begin{split}
\mathcal{L}_{\text{InfoNCE}}& = \\- \log &\frac{
\exp \left(sim(\mathbf{q}, \mathbf{k}^+)/\tau \right)
}{
\exp \left(sim(\mathbf{q}, \mathbf{k}^+)/\tau \right) + \sum\limits_{\mathbf{k}^- \in \mathcal{N}} \exp \left( sim(\mathbf{q}, \mathbf{k}^-)/\tau \right)
},
\end{split}
\end{equation}

\noindent
where \( \mathbf{q} \in \mathbb{R}^d \) denotes the query embedding vector obtained from a retrieved prompt.
 \( \mathbf{k}^+ \in \mathbb{R}^d \) denotes the positive key from the same class as the query \( \mathbf{q} \).
\( \mathcal{N} = \{ \mathbf{k}^-_1, \mathbf{k}^-_2, \dots, \mathbf{k}^-_M \} \) is a set of negative keys that do not correspond to \( \mathbf{q} \).
 \( \text{sim}(\mathbf{q}, \mathbf{k}) \) denotes a similarity function, typically cosine similarity, defined as \( \frac{\mathbf{q}^\top \mathbf{k}}{\|\mathbf{q}\| \|\mathbf{k}\|} \).
\( \tau > 0 \) is a temperature parameter that controls the sharpness of the distribution.
\sjcha{Finally, we train only the mapping network with $\mathcal{L}_{infoNCE}$ combined with Eq.\ref{eq-mse}, Eq.\ref{eq-rec}, and Eq.~\ref{eq-adv} losses as follows:} 

\begin{equation} \label{eq-total}
\begin{split}
\mathcal{L}_{total} = \lambda_{1}\mathcal{L}_{mse} + \lambda_{2}\mathcal{L}_{rec} + \lambda_{3}\mathcal{L}_{adv}+ \lambda_{4}\mathcal{L}_{infoNCE}.
\end{split}
\end{equation}
To balance the influence of different learning signals, we assign a separate weight $\lambda_{1}, \lambda_{2}, \lambda_{3}$ and $ \lambda_{4}$ to each loss term, which is determined empirically.

\section{Experiments}
\subsection{Experimental Setup}
\noindent\textbf{Training Configurations.} Our method is trained on the UrbanSound8K (US8K)~\cite{salamon2014dataset}, ESC-50~\cite{piczak2015esc}, and VGGSound~\cite{chen2020vggsound} datasets using a single NVIDIA RTX A6000 GPU.  
For VGGSound, following the experimental setup in Sound2Scene~\cite{lee2023generatingrealisticimagesinthewild}, we use a subset of 50 selected classes.
Images for evaluation are generated using Stable Diffusion v2.1~\cite{rombach2022high} with the PLMS sampler, and the random seed is fixed at 42 for reproducibility.
During training, we use a learning rate and batch size of $5 \times 10^{-5}$ and 8 for US8K, $1 \times 10^{-4}$ and 32 for ESC50, and $5 \times 10^{-4}$ and 256 for VGGSound.
The loss weights $\lambda_1$, $\lambda_2$, $\lambda_3$, and $\lambda_4$ are set to 1.0, 10000, 10000, and 0.5, respectively.  
For the InfoNCE loss, we use 8 negative samples $\mathcal{N}$ and set the temperature parameter $\tau$ to 0.8.

\noindent\textbf{Evaluation Metrics.}  
Since CLIP~\cite{radford2021learning} includes only text and image encoders, it is not inherently suitable for measuring the similarity between input audio and generated images.  
Following~\cite{yariv2023audiotokenadaptationtextconditioneddiffusion}, we adopt the Audio-Image Similarity (AIS) metric to evaluate cross-modal alignment between audio inputs and image outputs.  
% semantic space, which better assess cross-modal alignment.
In addition, we use Frechet Inception Distance (FID)~\cite{heusel2018ganstrainedtimescaleupdate} and Kernel Inception Distance (KID)~\cite{bińkowski2021demystifyingmmdgans} to assess the visual quality of the generated images.
Additional evaluation metrics, such as retrieval performance or user studies, are provided in the supplementary material.

\subsection{Comparisons with Previous Methods}
We compare CatchPhrase with Sound2Scene~\cite{sung2023sound}, SonicDiffusion~\cite{biner2024sonicdiffusion}, and GlueGen~\cite{qin2023gluegen} in Tab.~\ref{Comp_with_other_works}, all comparative experiments were conducted on the US8K, ESC50, and VGGSound50 datasets.
We report AIS scores only for US8K and ESC50, as these datasets contain audio-text pairs without corresponding images.
As shown in the table, our model consistently outperforms the baselines across all datasets.  
This result suggests that EXPrompts effectively help the generation model address both class-level and instance-level misalignment.

We also visualize qualitative results in Fig.~\ref{fig:Solving_Cross-modal_Ambiguity}. 
Unlike previous methods that often misalign the audio semantics, our model successfully generates images that match the corresponding audio inputs.  
In particular, CatchPhrase accurately generates the image for ``siren,'' compared to GlueGen, demonstrating its ability to resolve class-level ambiguity such as homographs.

\begin{table}[t]
\centering
\caption{ \label{Comp_with_other_works} \textbf{Quantitative comparison between existing audio to image generation methods and CatchPhrase on various audio classification dataset.} 
\textmd{Bolded numbers indicate the best performance, and underlined numbers represent the second-best scores across the compared methods.}}
%\begin{tabular*}{0.49\textwidth}{@{\extracolsep{\fill}}lccccc}
\resizebox{0.45\textwidth}{!}{
\begin{tabular}{lccccc}
    \hline			
    & \multicolumn{1}{c}{US8K}   & \multicolumn{1}{c}{ESC50}  & \multicolumn{3}{c}{VGGSound} \\
                    & AIS ($\uparrow$)           & AIS ($\uparrow$)            & AIS ($\uparrow$)            & FID ($\downarrow$)              & KID ($\downarrow$)  \\
    \hline  
    Sound2Scene~\cite{lee2023generatingrealisticimagesinthewild}     & .0842             & .1009              & .1080              & 87.04             & .0236 \\ 
    GlueGen~\cite{qin2023gluegen}         & \underline{.1444} & \underline{.1918}  &  \underline{.1950} & 76.66             & .0188             \\ 
    SonicDiffusion~\cite{biner2024sonicdiffusion}  & .0814             & .0624              & .0621              & \underline{68.98} & \underline{.0136} \\ 
    \rowcolor{bluei} CatchPhrase     & \bf{.1910}    & \bf{.2423}     & \bf{.2017}     & \bf{65.62}    & \bf{.0119}     \\ 
    \hline  
    		
%\end{tabular*}
\end{tabular}
}
\end{table}

\subsection{Ablation Studies}
\begin{figure}[t]
    \centering
    \includegraphics[width=0.9\linewidth]{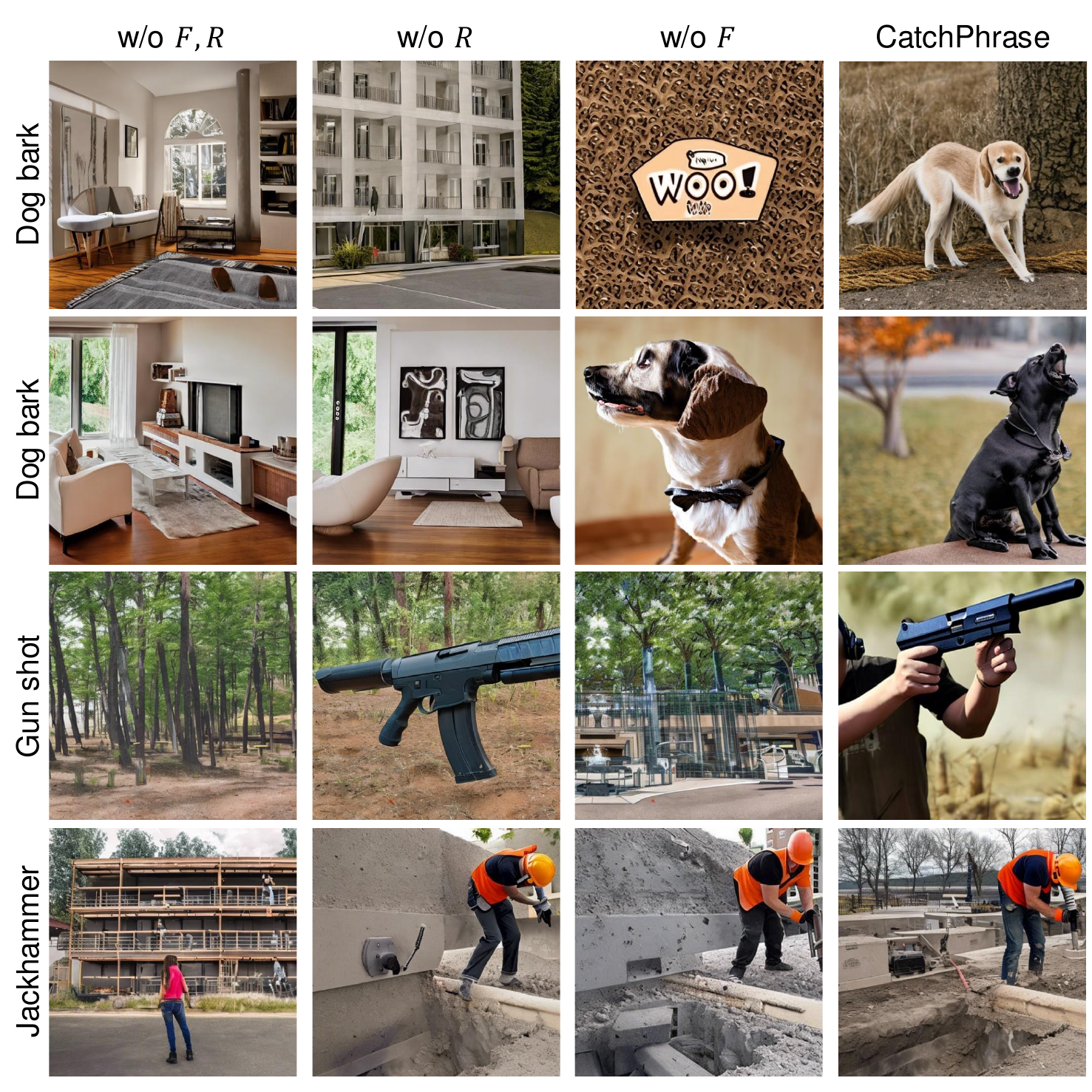} 
    \caption{\textbf{Qualitative results from the CatchPhrase component ablation study.} \textmd{Without applying filtering, prompts such as \textit{“woof woof”} may be semantically close but serve as noisy and cumbersome inputs from the perspective of image generation. Removing the retrieval component from CatchPhrase results in a loss of semantic precision during training.}
    }
    \label{fig:ablation-by-component}
\end{figure}

\noindent\textbf{Component Ablation Studies.}
%\textbf{Effectiveness of Each Component.}  
To evaluate the effectiveness of each component in CatchPhrase, we conduct a series of ablation studies comparing the template-based baseline with different configurations of our model. 
As shown in Tab.~\ref{ablation_component}, incorporating both filtering and retrieval leads to a clear improvement in audio-image alignment, demonstrating their complementary roles.  
Moreover, the qualitative results in Fig.~\ref{fig:ablation-by-component} further support this finding, illustrating that CatchPhrase produces more accurate and semantically aligned images compared to its ablated variants.
Overall, these results highlight the critical role of each component in achieving better performance.

% We also apply the retrieval module to GlueGen, which uses CLIP template-based prompts, and observe only marginal improvement.  

\begin{table}[t]
\centering
\caption{ \label{ablation_component} \textbf{Ablation studies for the CatchPhrase component.} \textmd{Assessing the impact of applying EXPrompt, filtering $F$ and retrieval $R$. We evaluate various dataset on AIS score. Higher is better.}}
%\begin{tabular*}{0.33\textwidth}{@{\extracolsep{\fill}}lcccc}
\resizebox{0.45\textwidth}{!}{
\begin{tabular}{lcccccc}
\hline
Method            &EXPrompt          &$F$           &$R$           & US8K & ESC50 & VGGSound\\ 
\hline
GlueGen~\cite{qin2023gluegen}             &                     &            &            & .1444 &.1918	&.1950\\

CatchPhrase         & \checkmark       &            &            & .0902 &.0876	&.0957\\

CatchPhrase         & \checkmark       & \checkmark &            & .1117 &.2005	&.1766\\

CatchPhrase         & \checkmark       &            & \checkmark & .1427 &.1570	&.1906\\

\rowcolor{bluei}CatchPhrase& \checkmark& \checkmark & \checkmark & \bf{.1910} &\bf{.2423}	&\bf{.2017}\\

\hline
%\end{tabular*}
\end{tabular}
}
\end{table}

\begin{table}[t]
\centering
\caption{ \label{table-query} \textbf{Ablation studies for queries.} \textmd{Class-based EXPrompt mining is conducted with different types of queries: $V$ (visual), $A$ (auditory), and $S$ (semantic).}}
\begin{tabular}{ccccc}%{0.5\textwidth}
%{@ccccc}%{@{\extracolsep{\fill}}ccccc}%{m{2cm}m{1cm}m{1cm}m{1cm}m{2cm}}%
\hline
Method       &$V$             &$A$              &$S$                & AIS ($\uparrow$)\\ 
\hline
GlueGen~\cite{qin2023gluegen}      &              &               &&  .1444\\
(a)          & \checkmark   &               &                 & .1671 \\

(b)          &              & \checkmark   &                 & .1529 \\
(c)          &              &              & \checkmark      & .1625 \\

(d)          & \checkmark   &               & \checkmark      & .1574\\

(e)          &              & \checkmark    & \checkmark      & .1590\\

\rowcolor{bluei}Ours        & \checkmark   & \checkmark    & \checkmark      &\textbf{.1910}\\
\hline
\end{tabular}

\end{table}

\noindent\textbf{Query Ablation Studies.} 
We investigate how different types of concept queries affect the final audio-visual alignment in Tab.~\ref{table-query}.  
As shown in the table, using all three query types—visual, auditory, and semantic—yields the best performance.  
A notable finding is that using only auditory queries leads to the lowest performance.  
This is because auditory prompts tend to produce rich prompts containing onomatopoeic expressions (e.g., ``woof''), which often lack semantic detail useful for image generation.  
Consequently, such prompts result in suboptimal outputs.
We provide examples generated using only auditory queries in the supplementary materials.

\begin{table}[t]
\centering
\caption{ \label{table-design} \textbf{Ablation studies for CatchPhrase design choices.} 
\textmd{We design four different variations of the CatchPhrase process, each of which is illustrated in the supplementary materials.}}
%\begin{tabular*}{0.39\textwidth}{@{\extracolsep{\fill}}lc}
\begin{tabular}{lc}%{0.39\textwidth}{@{\extracolsep{\fill}}lc}
\hline
Method                         & AIS ($\uparrow$)\\ 
\hline
CatchPhrase (Audio-only filtering)        & .1636 (+.0192)\\
CatchPhrase (Text-only filtering)        & .1724 (+.0280)\\
CatchPhrase (Merge after filtering)        & .1707 (+.0263)\\
\rowcolor{bluei}CatchPhrase (Merge before filtering)        & \textbf{.1910} (+.0466)\\
\hline
%\end{tabular*}
\end{tabular}
\end{table}

%from here....
\noindent\textbf{Filtering Equation Ablation Studies.}
As shown in Eq.~\ref{eq-filtering}, the filtering score consists of two main components:  
the first measures the similarity between a prompt and audio samples from the same class, 
while the second measures similarity to audio samples from different classes.
To evaluate the effectiveness of the second term, we conduct experiments on the US8K dataset using class-based EXPrompts.  
As presented in Tab.~\ref{table-filtering-equation}, incorporating the similarity term for different classes improves audio-image alignment.  
We attribute this to the fact that rich prompts occasionally contain details relevant to multiple classes, which may dilute class-specific semantics.  
By penalizing prompts that are overly similar to audio from other classes, the filtering mechanism effectively reduces cross-class semantic contamination.
\begin{table}[t]
\centering
\caption{ \label{table-filtering-equation} \textbf{Effect of penalizing cross-class similarity in relevance scoring.} \textmd{Penalizing prompts that are overly similar to audio samples from other classes helps filter out semantically ambiguous candidates, improving class-level disambiguation.}}
%\begin{tabular*}{0.36\textwidth}{@{\extracolsep{\fill}}lc}
\begin{tabular}{lc}
\hline
Method                  & AIS ($\uparrow$)\\ 
\hline
CatchPhrase (w/o negative term)     & .1652 (+.0208)\\

\rowcolor{bluei}CatchPhrase   & \textbf{.1910} (+.0466)\\

\hline
\end{tabular}

\end{table}

\noindent\textbf{Component Sequence Ablation Studies.} 
As shown in Fig.~\ref{fig:CatchPhrase-Method-Diagram}, CatchPhrase consists of three main stages: rich prompt generation, filtering, and retrieval.  
Each component can be flexibly configured depending on the modality and filtering strategy.

First, users may choose between using both audio and text inputs from the audio classification dataset or adopting a uni-modal approach.  
In the uni-modal setting, using only audio results in \textit{audio-only filtering}, while using only text leads to \textit{text-only filtering}.  
The filtering process involves two hyperparameters, $N_{as}$ and $N_{ft}$, which are selected empirically; corresponding experiments are included in the appendices.
After generating rich prompts, another design choice involves the point at which to apply filtering.  
In the \textit{merge-before-filtering} strategy, audio- and text-based prompts are first combined and then filtered.  
Conversely, \textit{merge-after-filtering} applies filtering to each modality independently before merging.

As shown in Tab.~\ref{table-design}, the merge-before-filtering strategy achieves superior performance.  
This improvement is attributed to the ability of text-based prompts to suppress noise introduced by audio-based prompts, which are more susceptible to auditory illusions caused by audio captioning models (ACMs).  
Filtering each modality separately may lead to cross-modal ambiguity, as noisy prompts from the audio stream can contaminate the final prompt set.  
In contrast, filtering after merging promotes the selection of semantically richer and more reliable prompts.

\begin{figure}[t]
    \centering
    \includegraphics[width=0.9\linewidth]{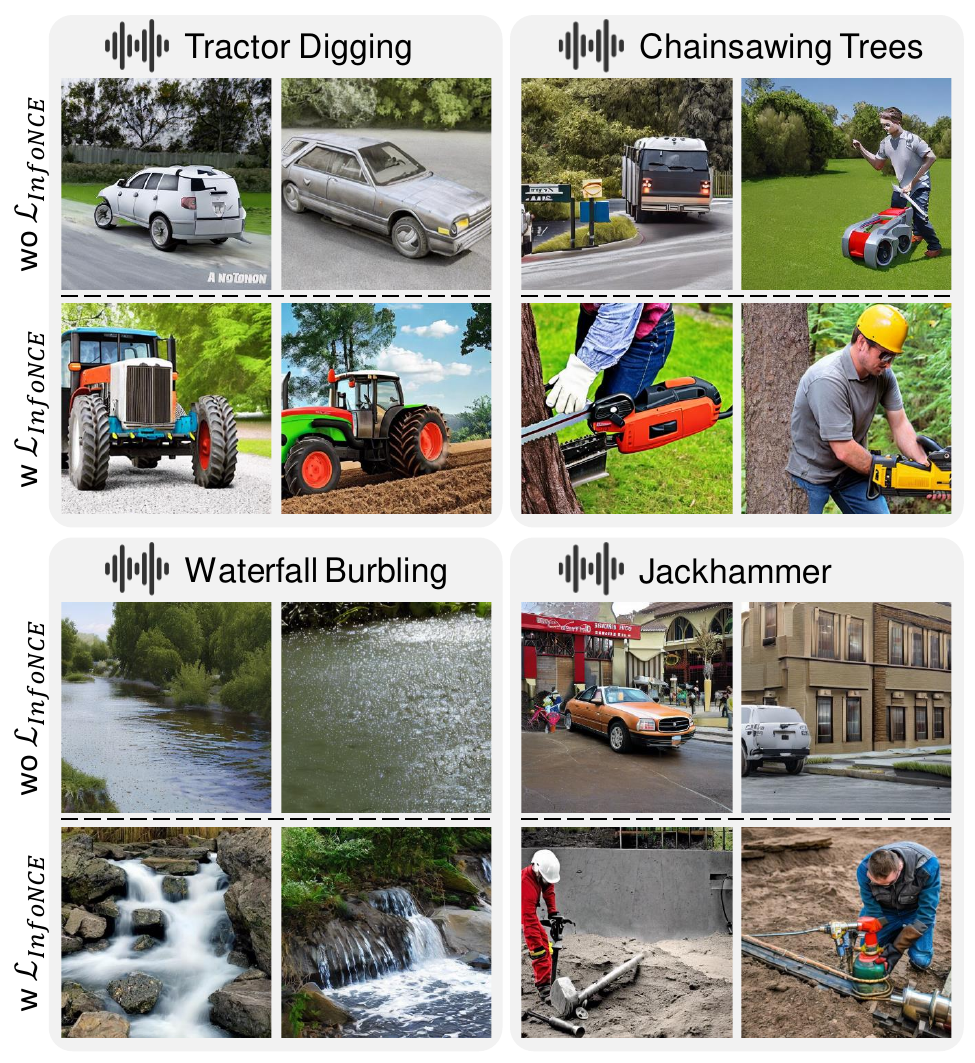} 
    \caption{\textbf{Comparison of image results with and without the use of InfoNCE loss.} \textmd{This shows the effectiveness of prompt-guided contrastive learning in enhancing semantic alignment.}
    }
    \label{fig:infonce}
\end{figure}

\noindent\textbf{Objective Function Ablation Studies.} 
Leveraging the filtered EXPrompts obtained through EXPrompt Selector for filtering within the InfoNCE loss contributes to mitigating auditory illusions. 
In contrast to the homograph issue, which occurs at the class level, auditory illusions arise at the instance level, making them more sensitive to sample-specific ambiguity.  
This effect is empirically validated in Tab.~\ref{table-infoNCE} and Fig.~\ref{fig:infonce}.
\begin{table}[t]
\centering
\caption{ \label{table-infoNCE} \textbf{The effect of leveraging InfoNCE objective.} \textmd{We evaluate various dataset on AIS score. Higher is better.}}
\begin{tabular}{lccc}
%\begin{tabular*}{0.43\textwidth}{@{\extracolsep{\fill}}lccc}
\hline
Method                  & US8K &ESC50 &VGGSound\\ 
\hline
CatchPhrase (w/o InfoNCE)     & .1905 &.2422 &.2010\\
\rowcolor{bluei}CatchPhrase (Ours)   & \textbf{.1910} 
 &\textbf{.2423} &\textbf{.2017}\\
\hline
%\end{tabular*}
\end{tabular}
\end{table}

\noindent\textbf{Audio Captioning Models with Image Generation.}
Audio-to-image generation using audio captioning models (ACMs) can be approached in several ways.  
First, captions for all audio samples in an audio classification dataset can be pre-generated using ACMs and used for one-to-one supervised training.  
Second, the pre-generated captions can serve as direct inputs to a text-to-image generation model, enabling a two-stage audio-to-image generation pipeline.
We compare these two approaches with CatchPhrase on the US8K dataset.  
As reported in Tab.~\ref{table-AudioCapting}, CatchPhrase outperforms both ACM-based methods in terms of AIS scores.
These results suggest that the inferior performance of ACM-based approaches stems from inaccurate or ambiguous captions, likely caused by auditory illusions during audio-only filtering.
They also demonstrate that directly using such captions as prompts in a text-to-image generation model is less effective than training an audio-to-image model that learns the underlying semantics of the audio input.

\begin{table}[t]
\centering
\caption{\label{table-AudioCapting} \textbf{Ablation on auido caption to image generation.}\textmd{We evaluate various frameworks that can use text descriptions extracted from audio inputs.} }
%\begin{tabular*}{0.29\textwidth}{@{\extracolsep{\fill}}lccc}
\begin{tabular}{lc}
\hline
Method                  & AIS ($\uparrow$)\\ 
\hline

GlueGen~\cite{qin2023gluegen}   & .1444\\

GlueGen (w/ Audio Caption)      & .1307\\
EnCLAP~\cite{kim2024enclap}+Stable Diffusion2.1~\cite{rombach2022high} &.1695\\
\rowcolor{bluei} CatchPhrase        & \textbf{.1910}\\

\hline
%\end{tabular*}
\end{tabular}

\end{table}

\begin{figure}[t]
    \centering
    \includegraphics[width=0.9\linewidth]{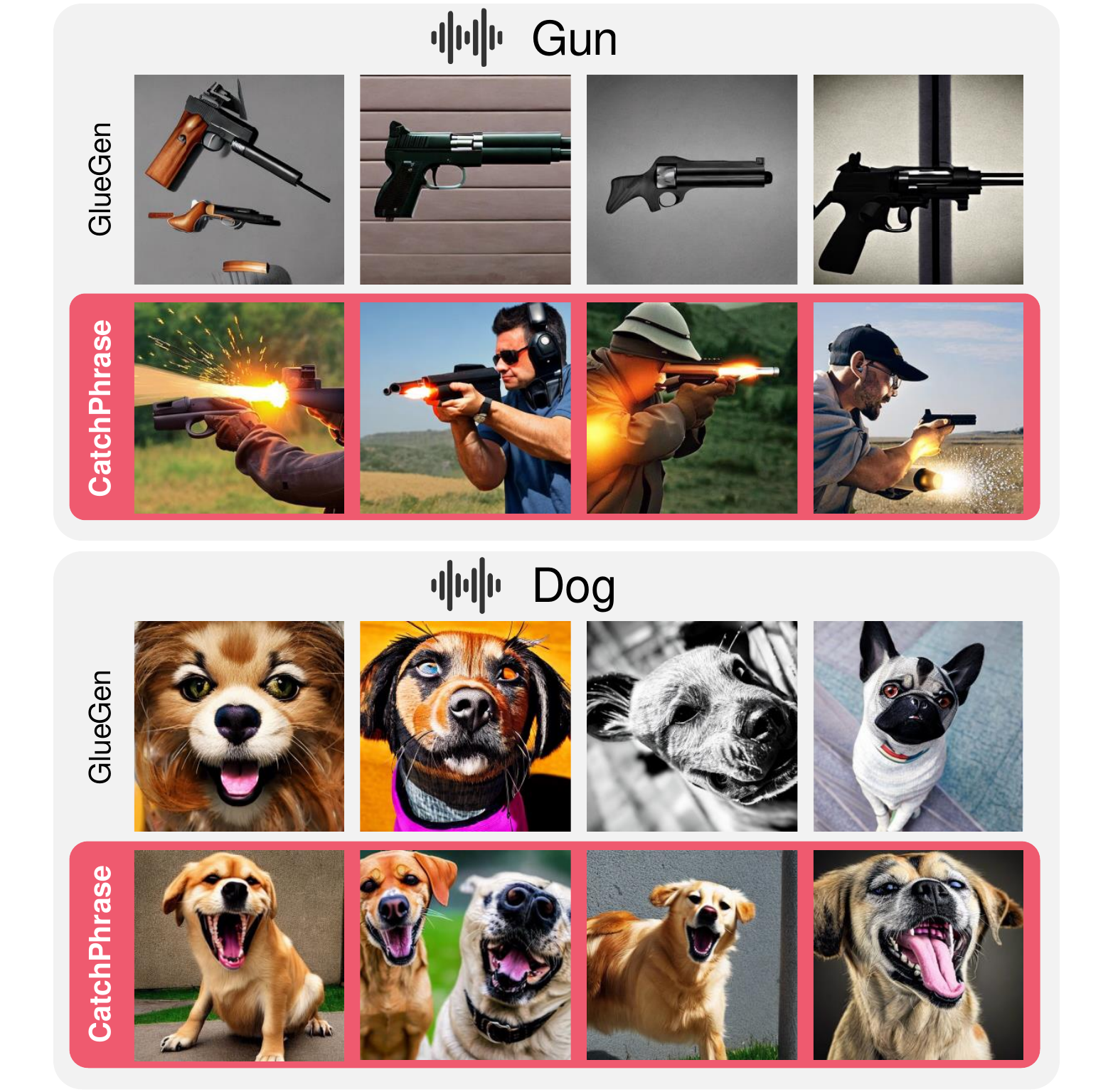}
    \caption{\textbf{Impact of EXPrompts on image generation compared to template-based prompts.}
    \textmd{Images with a white borderline are generated using template-based prompts, while images with a pink borderline are generated using CatchPhrase.}
    }
    \label{fig:rich_expression}
\end{figure}

\noindent\textbf{Effectiveness of EXPrompts.}
We investigate the impact of rich prompts on image generation compared to template-based prompts in Fig.~\ref{fig:rich_expression}.
The ESC50 dataset contains simpler class labels than US8K or VGGSound, with sounds such as ``gunshot'' or ``dog bark'' labeled simply as ``gun'' or ``dog.''  

As can be seen, template-based prompts often fail to capture the full context of the audio input, leading to less accurate visual representations.  
In contrast, CatchPhrase generates rich, context-aware descriptions based on both audio and text inputs, enabling more faithful image generation.  
For example, even when the audio input corresponds to the word ``gun,'' CatchPhrase generates an image of a gun emitting sparks, effectively reflecting the sound of a gunshot.
Similarly, for the class ``dog,'' it generates an image of a barking dog that aligns with the audio input.
This demonstrates that EXPrompts enrich the semantic information of weak labels in conventional audio classification datasets, thereby improving semantic alignment between audio and text.

\section{Conclusion}
% Our proposed method, CatchPhrase, addresses cross-modal ambiguity arising from weak labels and audios in audio classification datasets for audio-to-image generation. 
% By harnessing LLMs and ACMs, we generate a diverse set of enriched prompts while eliminating the need for predefined templates that entail additional human effort. 
% To mitigate residual ambiguity, we introduce a multimodal-aware filtering mechanism that selects the most suitable prompts for training. 
% Experiments across multiple datasets validate the efficacy of our approach, demonstrating superior performance in both quantitative and qualitative assessments.
We propose CatchPhrase, a framework designed to address cross-modal misalignment in audio-to-image generation, particularly those stemming from weak labels and ambiguous audio in audio classification datasets.  
By leveraging large language models (LLMs) and audio captioning models (ACMs), our method generates a diverse set of enriched prompts without relying on predefined templates, thereby reducing manual effort.  
To further alleviate residual misalignment, we introduce a multimodal-aware filtering mechanism that selects semantically appropriate prompts for training the encoder adaptor.  
Extensive experiments on multiple datasets validate the effectiveness of our approach, demonstrating consistent improvements in both quantitative metrics and qualitative results.

\begin{acks}
This was partly supported by the Institute of Information \& Communications Technology Planning \& Evaluation (IITP) grant funded by the Korean government(MSIT) (No.RS-2020-II201373, Artificial Intelligence Graduate School Program(Hanyang University)) and the Institute of Information \& Communications Technology Planning \& Evaluation (IITP) grant funded by the Korean government(MSIT) (No.RS-2025-02219062, Self-training framework for VLM-based defect detection and explanation model in manufacturing process).
\end{acks}

\bibliographystyle{ACM-Reference-Format}
\bibliography{sample-base}

\clearpage
%\newpage
\appendix
%\onecolumn
\begin{center}
\Huge \textbf{Supplementary Materials}
\end{center}
Additional experimental details and analyses, which complement the main paper but were excluded due to space limitations, are provided in this supplementary material.

\section{Comparison of Image Generation Results Based on the Employed Queries}
\begin{figure}[h!]
    \centering
    \includegraphics[width=0.9\linewidth]{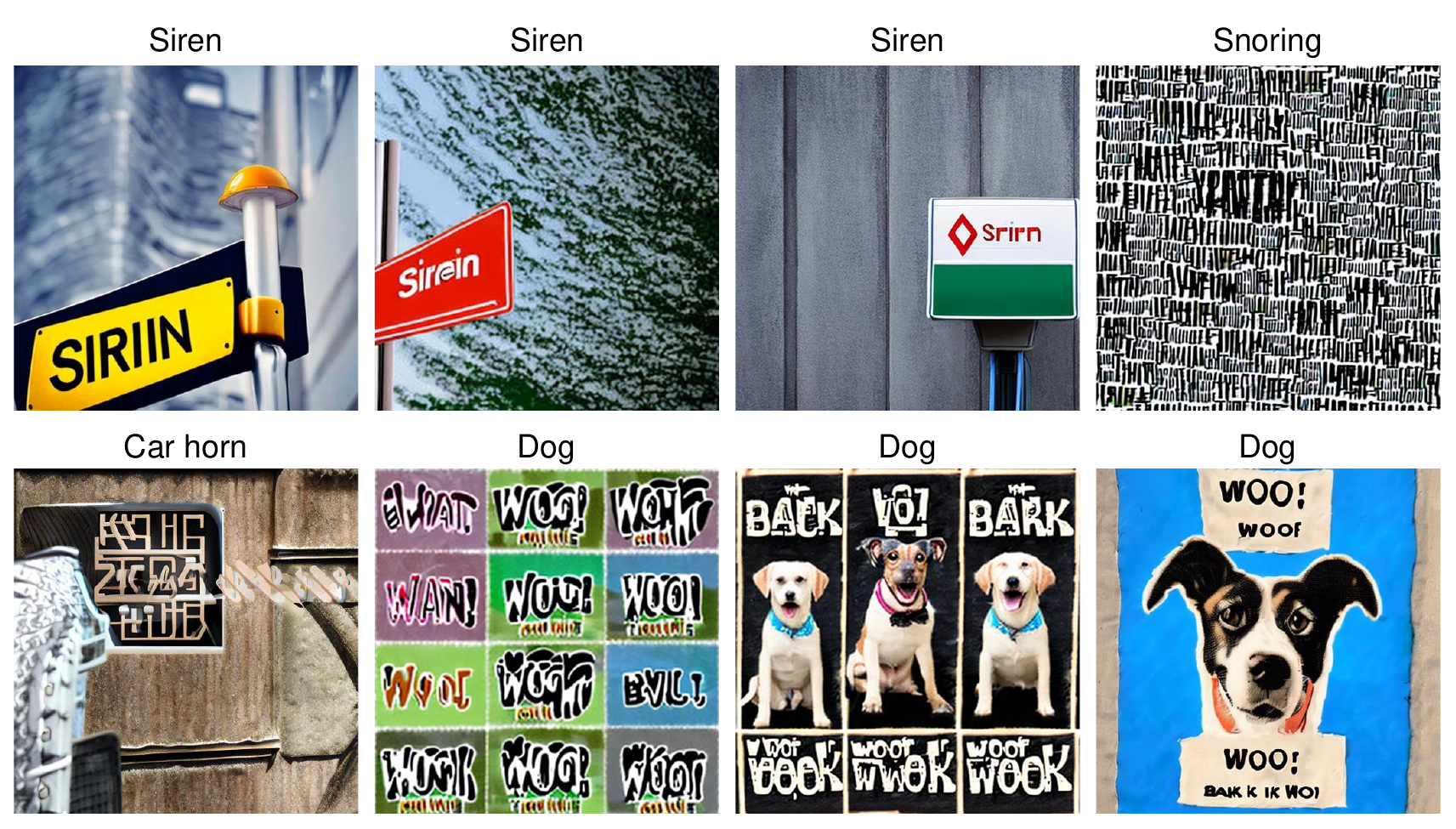} 
    \caption{\textbf{Generated images with using audio queries in text-based rich prompts.} \textmd{Onomatopoeic expressions and textual elements observed in images generated using EXPrompts crafted from audio queries, among text-based EXPrompts.}}
    \label{fig:audio-queries}
\end{figure}

In experiment section, the use of audio queries is associated with a reduction in AIS performance.
The observed quality degradation when utilizing only audio queries can be attributed to the model’s tendency to prioritize expressing the auditory characteristics of the given class label rather than incorporating visual information to guide image generation. 
As a result, the model often generates images that depict onomatopoeic expressions rather than faithfully constructing the intended objects, as illustrated in Fig.~\ref{fig:audio-queries}. 

\section{EXPrompts with Paraphrasing}
After filtering the EXPrompts, we use paraphrasing to augment the number of valuable prompts in \textit{text-only filtering}. Paraphrasing leverages LLMs with the following query, similar to generating rich expression prompts. 
This process generates augmented prompts five times more than the number of filtered prompts used in retrieval along with the original filtered prompts.% ~\cite{abaskohi2023lm}. 
Filtering applies again, and related experiments reveal that paraphrasing does not significantly enhance performance. 
This outcome results from paraphrasing, altering only the style of sentences while maintaining the semantics of the already filtered prompts. 
This leads to minimal differences in outcomes, as shown in Tab~\ref{table-paraphrasing}. 
For comparison, CatchPhrase uses filtering hyper-parameters that sample 10 audio samples per class and select the top 10 prompts. 
The queries used for prompt augmentation are as follows: 
\begin{mybox}\texttt
\begin{itemize}
\item Summarize the following text in your own words: \textit{prompt} 
\item Rewrite the following text that expresses the same idea in a different way: \textit{prompt} 
\item Generate a paraphrase of the following text that expresses the same ideas in a different way: \textit{prompt}
\item Generate a paraphrase of the following text using different words and sentence structures \\while still conveying the same meaning: \textit{prompt}
\item Generate a summary or paraphrase of the following text that captures the essence of the ideas \\in a concise manner: \textit{prompt}
\end{itemize}
\end{mybox}

\begin{table}[t]
\centering
\caption{ \label{table-paraphrasing} \textbf{Quantitative comparison for applying paraphrase.} \textmd{$P$ denotes paraphrasing, and $F$ denotes filtering. The numbers in parentheses indicate the difference between AIS and GlueGen.}}
%\begin{tabular*}{0.44\textwidth}{@{\extracolsep{\fill}}lccc}
\begin{tabular}{lccc}
\hline
Method                 &P   &F & AIS\\ 
\hline
CatchPhrase (w/ $P$)   &\checkmark & & .1403 (-.0041)\\

CatchPhrase (w/ $P,F$) &\checkmark &\checkmark  & .1584 (+.0140)\\

\rowcolor{bluei}CatchPhrase (Text-only filtering) & &\checkmark  & \textbf{.1724 (+.0280)}\\
\hline
\end{tabular}
\end{table}

\begin{figure*}[t]
    \centering
    \includegraphics[width=0.7\linewidth]{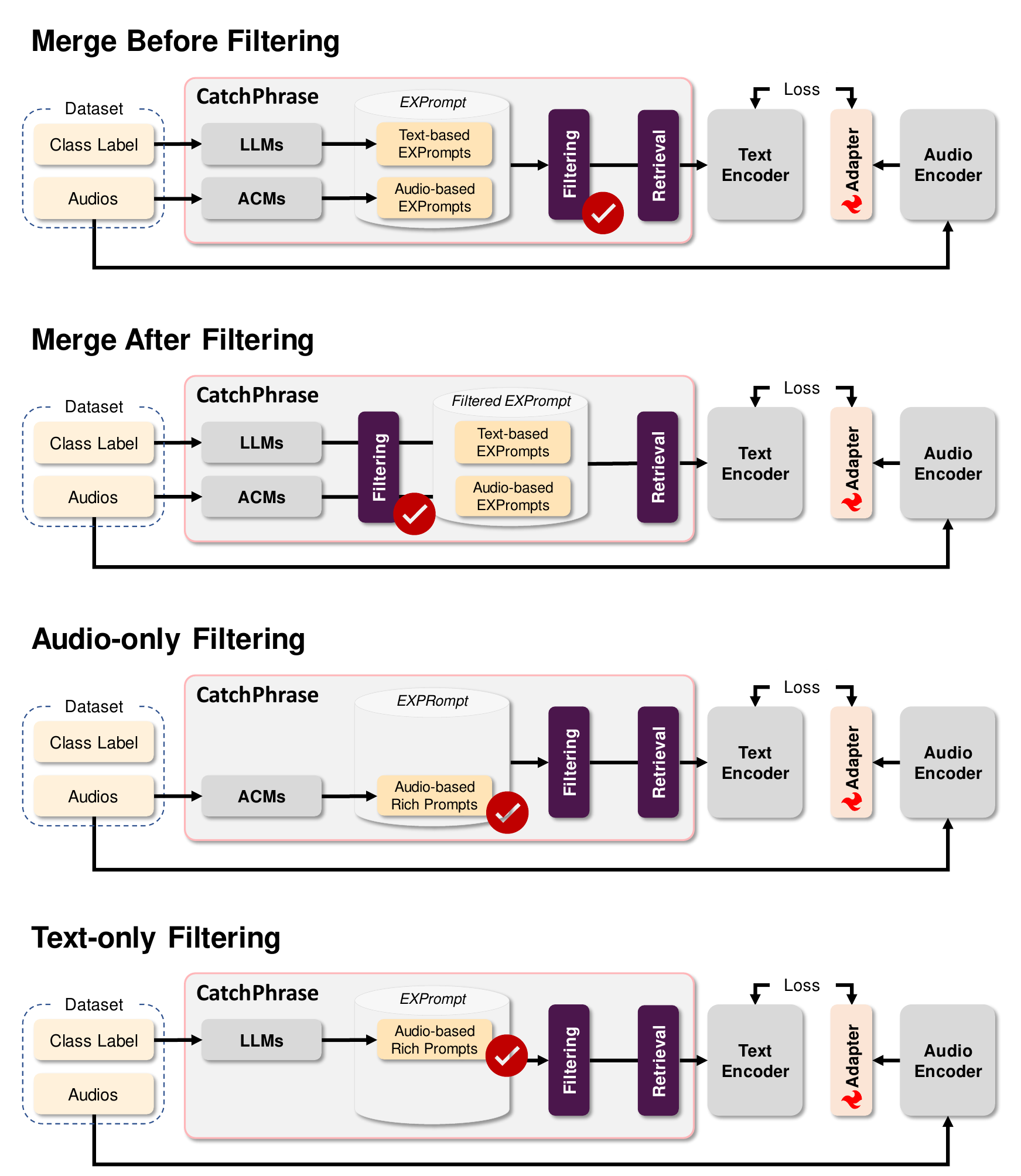} 
    \caption{\textbf{Various component sequences in consisting CatchPhrase.}}
    \label{fig:design-choices}
\end{figure*}
\section{Component Sequence in CatchPhrase}
The components of CatchPhrase can be configured in various combinations.
To evaluate the effectiveness of each configuration, we measured the AIS for possible combinations.
To facilitate a clearer understanding of these configurations, we have included Fig.~\ref{fig:design-choices} in the supplementary material.
The \textit{Merge before filtering} represents our most effective component configuration.

\section{Filtering Parameter}
Among the filtering parameters, the choice of top-K yields intriguing experimental results as shown in Tab.~\ref{table-clap-based-filtering}. 
The top row shows the number of audio samples selected per class ($N_{AS}$) when constructing the audio subset. 
The first column shows the cutoff value for EXPrompts. The table values represent AIS scores. 
The middle row displays the AIS for text-based rich prompt filtering parameters, while the bottom row shows the AIS for audio-based rich prompt filtering parameters.

\begin{table*}[t]
\centering
\caption{ \label{table-clap-based-filtering} \textbf{Experiment for CLAP-based filtering parameters.}  
\textmd{\textit{cls} denotes the abbreviation of class. For US8K and VGGSound, audio subsets of 10, 50, and 100 samples per class were constructed, whereas for ESC-50, subsets of 10, 20, and 30 samples per class were used due to its smaller class size. Performance was measured and compared using the AIS score (higher is better)}}
\begin{tabular*}{0.7\textwidth}{@{\extracolsep{\fill}}l|ccc|ccc|ccc}
    \hline			
    & \multicolumn{3}{c|}{US8K}   & \multicolumn{3}{c|}{ESC50}  & \multicolumn{3}{c}{VGGSound} \\
             & 10/cls   & 50/cls           & 100/cls    & 10/cls             & 20/cls    & 30/cls   & 10/cls   & 50/cls           & 100/cls \\
    \hline  
    Top-1    & .1346    & .1388            & .1400      & .2155     & .2246     & \textbf{.2304}    & .1966    & .1914   & \textbf{.2004} \\ 
    Top-5    & .1594    & \textbf{.1724}   & .1619      & .1538              & .1513     & .1663    & .1534    & .1665   & .1602  \\
    Top-10   & .1630    & .1572            & .1659      & .1389              & .1545     & .1420    & .1453    & .1392   & .1476 \\
    Top-20   & .1559    & .1538            & .1648      & .1093              & .1193     & .1229    & .1374    & .1427   & .1384 \\
    \hline  
    Top-1    & .1114            & .1091    & .1075  & .1718     & .1763             & .1827    & .1036    & .1127   & .1136 \\ 
    Top-5    & .1380            & .1316    & .1347  & .1934     &\textbf{.2035}     & .1982    & .1288    & .1287   & \textbf{.1313}  \\
    Top-10   & .1457            & .1584    & .1511  & .1994     & .1981             & .1847    & .1264    & .1249   & .1235 \\
    Top-20   & \textbf{.1636}   & .1574    & .1588  & .1993     & .1815             & .1953    & .1213    & .1277   & .1282 \\
    \hline				
\end{tabular*}
\end{table*}

\begin{figure}[t]
    \centering
    \includegraphics[width=1\linewidth]{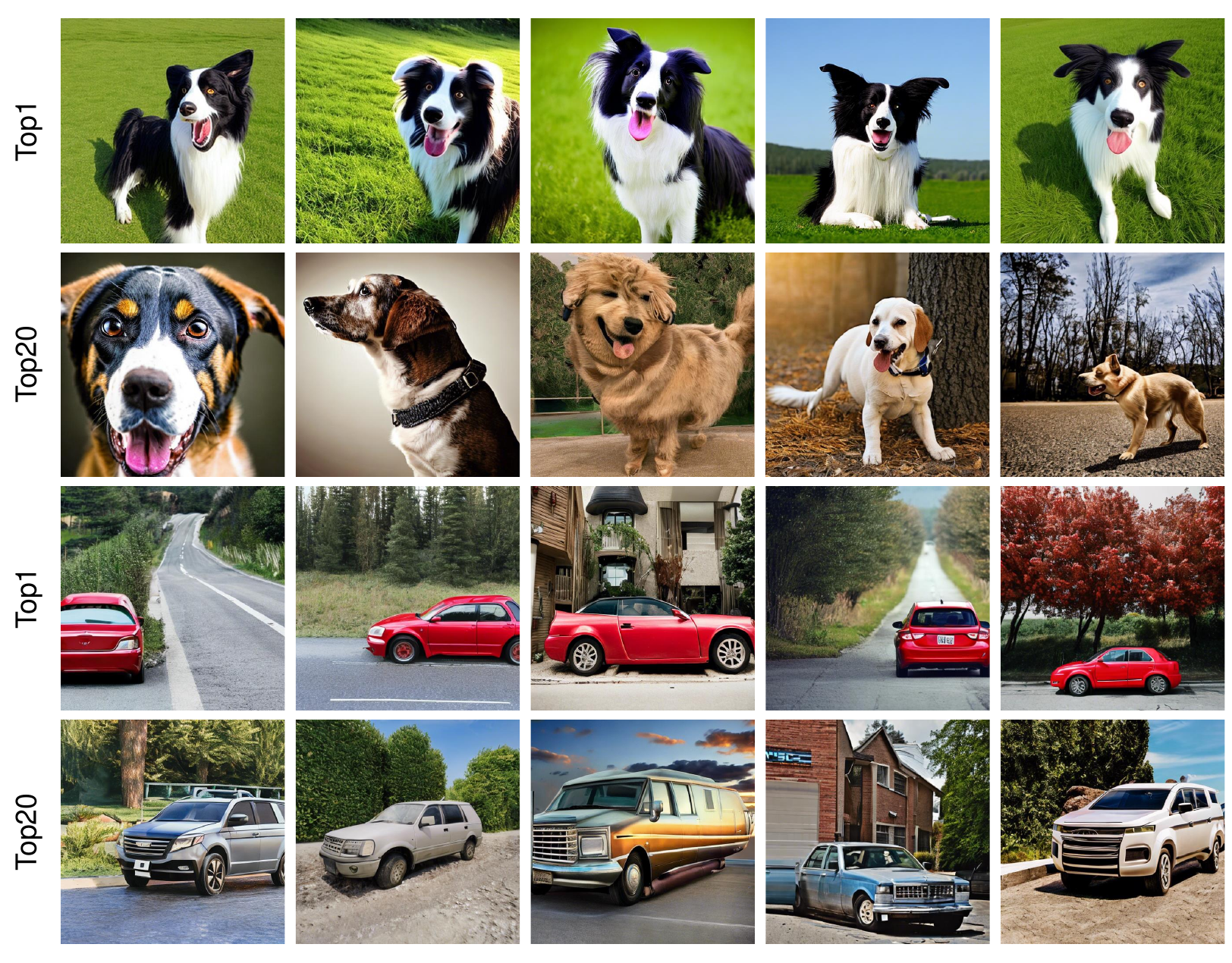} 
    \caption{\textbf{Differences in generated images based on the top-K parameter in the filtering method.}
    \textmd{A counter effect is observed during filtering: as the value of top-K decreases, the diversity of the generated images also decreases.}}
    \label{fig:ablation-topk}
\end{figure}

When different audio inputs from the same class are processed, a smaller value of $N_{EXP}$ leads to a noticeable reduction in the diversity of the generated images. 
As illustrated in Fig.~\ref{fig:ablation-topk}, a clear difference in diversity can be observed between the cases where $N_{EXP}=1$ and $N_{EXP}=20$ for audio inputs belonging to the ``dog'' and ``engine idling'' classes.

\section{More on EXPrompt Mining}
In this section, we present the exact queries used for EXPrompt construction (see Tab.~\ref{table-quiries}) and instance-based prompts (see Tab.~\ref{table-class-based EXPrompt}) and instance-based EXPrompts (see Tab.~\ref{table-instance-based EXPrompt}), corresponding to class-level and instance-level semantics, respectively.

\begin{table*}[t]
\centering
\caption{ \label{table-quiries} \textbf{Set of queries employed for EXPrompt generation, conditioned on class labels.}\textmd{ Visual, Auditory, and Semantic queries through large language models. The bolded words represent the key terms in each query.}}
\begin{tabular}{m{2cm}m{15cm}}%{1\textwidth}{@{\extracolsep{\fill}}m{5cm}m}
\hline
Category                & Queries\\ 
\hline
           & Describe what a(n) \_\_\_ \textbf{\textit{looks}} like in real world. \\
Visual     & What does a(n) \_\_\_ \textbf{\textit{looks}} like in real world?\\
           & Describe an image from the internet of a(n) \_\_\_ \textbf{\textit{looks}} in real world.\\
\hline
           & Describe what a(n) \_\_\_ \textbf{\textit{sounds}} like in real world. \\
Audio      & What does a(n) \_\_\_ \textbf{\textit{sound}} like in real world?\\
           & Describe a \textbf{\textit{sound}} from the internet of a(n) \_\_\_ in real world.\\
\hline
           & Create one sentence about \textbf{\textit{meaning}} of a(n) \_\_\_ in real world: \\
Semantic   & \textbf{\textit{Summarize}} a(n) \_\_\_ in a single sentence.\\
           & Describe what a(n) \_\_\_ \textbf{\textit{represents}} in a real-world context in one sentence. \\
\hline
\end{tabular}

\end{table*}

\begin{table*}[t]
\centering
\caption{ \label{table-class-based EXPrompt} \textbf{Example of class-based EXPrompt.}\textmd{ Representative EXPrompts generated by processing weak class labels with visual, auditory, and semantic queries through large language models. The bolded words indicate either information related to other classes or the presence of homographs within each EXPrompt.}}
\begin{tabular}{m{2cm}m{15cm}}%{1\textwidth}{@{\extracolsep{\fill}}m{5cm}m}
\hline
Class                & Generated Rich Prompts\\ 
\hline
           & A cricket is a small \textbf{insect} that typically measures around  inch in length It has a round and flattened body.\\
           & Crickets are small, insectivorous \textbf{insects} that are commonly found in grassy areas.\\
Crickets   & The image shows a close-up of a group of crickets crawling on a patch of green grass.\\
           & A cricket's sound is often described as a high-pitched chirping\\
           & In the real world, crickets make a chirping or buzzing sound.\\
\hline
           & The constant buzzing of wings as tiny insects dart through the air.\\
           & The sound of \textbf{crickets} chirping loudly can be heard in the background.\\
Insects    & The buzzing symphony of cicadas, \textbf{crickets}, and grasshoppers fill the warm summer air.\\
           & The buzzing of a thousand tiny wings fills the air.\\
           & An insect buzzing.\\
\hline
           & A siren is a type of loud noise-making device used for emergency purposes.\\
           & A \textbf{mythological creature} with the upper body of a beautiful woman and the lower body of a bird or fish.\\
Siren      & A \textbf{beautiful, ethereal siren} perched on a rocky outcrop overlooking the ocean.\\
           & A siren in the real world sounds like a loud, high-pitched wailing noise.\\
           & A siren is a loud warning device used to alert people of impending danger.\\
\hline
\end{tabular}

\end{table*}

\begin{table*}[h!]
\centering
\caption{ \label{table-instance-based EXPrompt} \textbf{Example of instance-based EXPrompt.} \textmd{Representative EXPrompts generated by audio instance through ACMs. The bolded words indicate the presence of information related to other classes within each EXPrompt.}}

\begin{tabular}{m{2cm}m{15cm}}%{1\textwidth}{@{\extracolsep{\fill}}ll}
\hline
Class      & EXPrompts\\ 
\hline
           & Wind blows and \textbf{birds chirp} in the distance.\\
           & Crickets chirp and \textbf{frogs croak}.\\
Crickets   & A small \textbf{motor is running and rattling}.\\
           & Frogs croaking and \textbf{birds chirping}.\\
           & Crickets chirp in the distance.\\
\hline
           & \textbf{A man speaking} followed by rustling and buzzing\\
           & Crickets chirp and insects buzz.\\
Insects    & Insects buzzing and \textbf{birds chirping}.\\
           & \textbf{Birds are chirping} and insects are buzzing.\\
           & Insects are buzzing and \textbf{birds are chirping}.\\
\hline
           & An emergency siren is triggered and passes by.\\
           & An emergency siren is triggered and moves closer.\\
Siren      & An emergency siren is triggered and moves closer.\\
           & An emergency siren is triggered and a horn is triggered.\\
           & An emergency siren is triggered while \textbf{people talk} in the background.\\
\hline
\end{tabular}
\end{table*}

\section{Audio to Text Retrieval}
To evaluate whether EXPrompt is truly related to the audio, we compared it against the predefined template-based prompts used in the original GlueGen. 
Specifically, we measured whether the prompt with the highest retrieval top-1 similarity to the test set audio samples originated from EXPrompt or from the baseline prompts. 
For this evaluation, we employed CLAP as the text and audio encoder. 
The results show that the proposed EXPrompt achieves higher audio-to-text retrieval performance compared to the baseline method, indicating that our approach constructs text representations that are more closely aligned with the audio content.

\begin{table}[t]
\centering
\caption{ \label{table-a2tR} \textbf{Audio to Text retrieval results.} \textmd{The retrieval results are reported in terms of R@1.}}
%\begin{tabular*}{0.44\textwidth}{@{\extracolsep{\fill}}lccc}
\begin{tabular}{lccc}
\hline
Method           &US8K &ESC50 &VGGSound\\ 
\hline
GlueGen~\cite{qin2023gluegen}   &11.3\%	&25.5\%	&10.4\%\\
\rowcolor{bluei}CatchPhrase &\bf{88.7\%}	&\bf{74.5\%}	&\bf{89.6\%}\\
\hline
\end{tabular}
\end{table}

\section{User Studies}
A total of 30 participants were asked to compare 15 pairs of images generated from audio inputs using GlueGen and CatchPhrase. 
Participants responded to two questions: (1) ``Which image is most relevant to the given audio?'' to evaluate the perceptual alignment between the audio and the generated images, and (2) ``Which image is most relevant to the given audio and its associated text?'' to assess whether additional textual information helps mitigate auditory illusion effects. 
As shown in Tab.~\ref{table-userstudies} suggest that CatchPhrase more effectively aligns audio and image semantics, especially in cases where textual context is required to resolve ambiguity. This supports the claim that our method successfully incorporates enriched prompts to improve cross-modal understanding.

\begin{table}[t]
\centering
\caption{ \label{table-userstudies} \textbf{User studies results.} \textmd{(1), (2) denote the questions asked to participants.}}
%\begin{tabular*}{0.44\textwidth}{@{\extracolsep{\fill}}lccc}
\begin{tabular}{lcc}
\hline
Method                 &(1)   &(2) \\ 
\hline
GlueGen~\cite{qin2023gluegen}   &35.78\%	&19.78\%
\\
\rowcolor{bluei}CatchPhrase &\bf{64.22\%}	&\bf{80.22\%}\\
\hline
\end{tabular}
\end{table}

\section{Qualitative Results for EXPrompt Scoring}
\begin{figure}[t]
    \centering
    \includegraphics[width=1\linewidth]{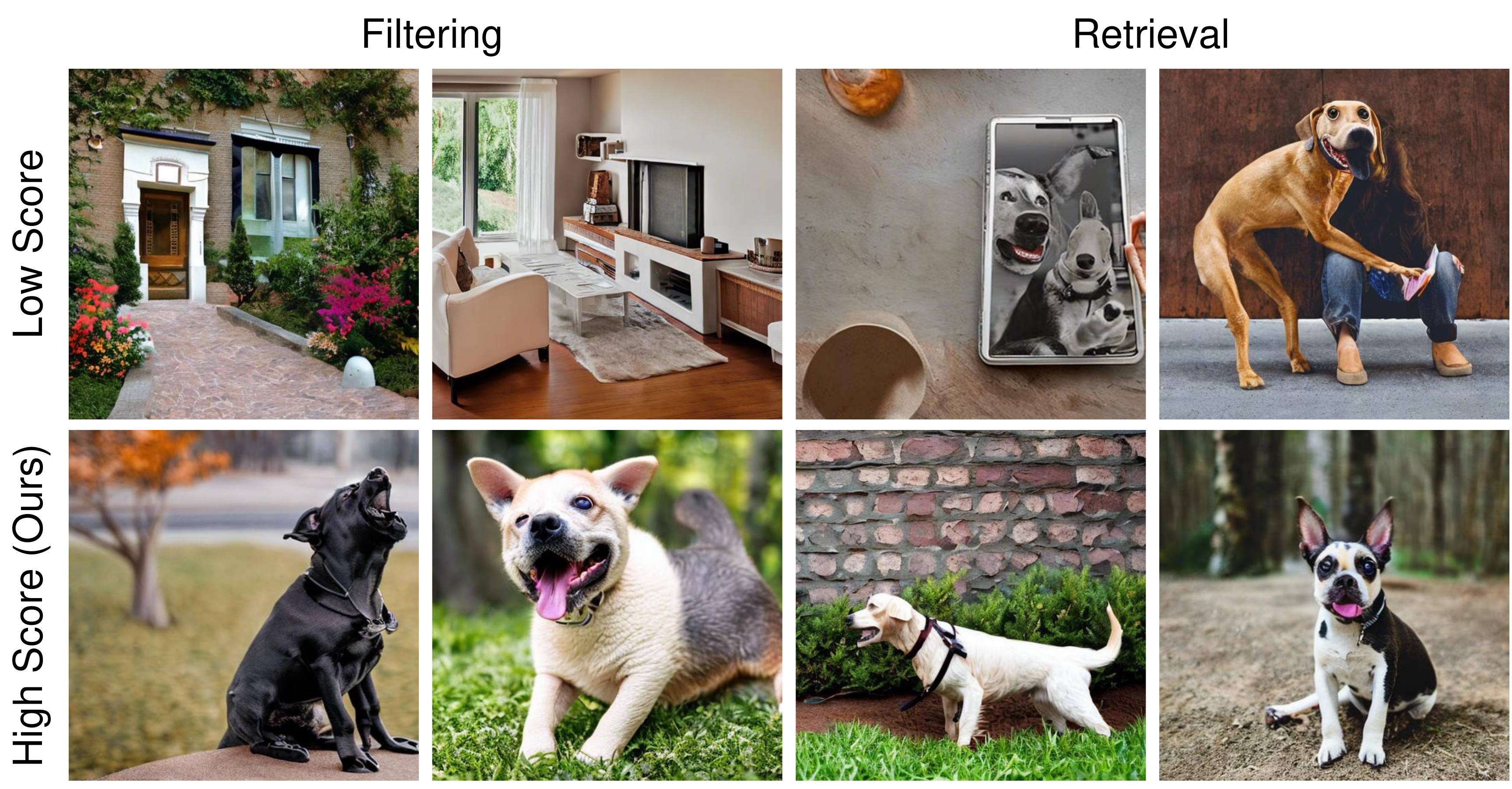} 
    \caption{\textbf{Qualitative results for low- vs. high-scoring EXPrompts in filtering and retrieval.}
    \textmd{The audio samples used in this experiment are dog bark recordings from the US8K dataset. The top row shows images generated after training with low-scoring prompts, while the bottom row shows results using high-scoring EXPrompts selected by our proposed method. Since the scoring is applied at both the filtering and retrieval stages, we present comparisons for each component.}}
    \label{fig:high_low_score-topk}
\end{figure}

We applied the proposed scoring method across both filtering and retrieval components. 
While the individual contributions of each component were validated through ablation studies, we additionally examined qualitative results to assess how high- versus low-scoring EXPrompts affect the generated images. 
As shown in Fig.~\ref{fig:high_low_score-topk}, high-scoring prompts lead to images that are semantically aligned with the corresponding audio, whereas low-scoring prompts tend to produce visually plausible but semantically mismatched outputs. This observation underscores the effectiveness of our scoring and selection strategy.

Notably, in the filtering stage, the presence of noisy prompts generated by LLMs and ACMs causes low-scoring EXPrompts to yield largely random images. In the retrieval stage, although the EXPrompts have already been filtered once, using low-scoring prompts still results in images that reflect only the dominant class information (e.g., ``dog'') from the audio, failing to capture finer details such as the sub-concept ``bark'' and degrading overall image quality. By contrast, our approach—applying high-scoring EXPrompts in both filtering and retrieval—successfully preserves the semantic content of the audio and enhances the fidelity of the generated images.

\section{Limitation}
Despite the promising results, our study has several limitations that warrant further investigation. 
First, due to the limited information available in the training data, zero-shot audio-to-image generation remains an open and challenging problem that has yet to be thoroughly explored. 
Second, generating images from audio with multiple overlapping events continues to pose significant challenges, as disentangling and accurately representing such complex auditory scenes is inherently difficult. 
Lastly, incorporating temporal segmentation could potentially improve the model’s ability to handle sequential events in audio. 
Addressing these challenges constitute an important direction for future research.

\end{document}